\documentclass[galaxies,review,abstract,accept,moreauthors]{Definitions/mdpi}

\firstpage{1} 
\pubvolume{13}
\issuenum{4}
\articlenumber{82}
\pubyear{2024}
\copyrightyear{2024}
\externaleditor{Dominic Bowman}
\datereceived{26 February 2025} 
\daterevised{9 June 2025} 
\dateaccepted{17 June 2025} 
\datepublished{21 July 2025} 
\hreflink{https://doi.org/10.3390/galaxies13040082} 

\newcommand{\s}{\textsuperscript}
\usepackage{xcolor}

\Title{Molecules and Chemistry in Red Supergiants}

\TitleCitation{Molecules and Chemistry in Red Supergiants}

  \Author{{Lucy M. Ziurys} 
 $^{1,}$* and
  Anita M. S. Richards $^{2,}$*\orcidB{}}

\AuthorNames{Lucy M. Ziurys and   Anita M. S. Richards}

\AuthorCitation{{Ziurys, L.M.;} 
 Richards, A.M.S.} 

\address{%
$^{1}$ \quad {Department of Astronomy,  Department of Chemistry, Arizona Radio Observatory,} 
 University of Arizona, Tucson, {AZ  85721-0065}
, USA\\
$^{2}$ \quad {JBCA,} 
  Department Physics and Astronomy, University of Manchester, Manchester M13 9PL, UK}

\corres{Correspondence:  lziurys@arizona.edu (L.M.Z.); a.m.s.richards@manchester.ac.uk (A.M.S.R.)}

\abstract{The envelopes of Red Supergiants (RSGs) have a unique chemical environment not seen in other types of stars. They foster an oxygen-rich synthesis but are tempered by sporadic and chaotic mass loss, which distorts the envelope and creates complex outflow sub-structures consisting of knots, clumps, and arcs. Near the stellar photosphere, molecules and grains form under approximate LTE conditions, as predicted by chemical models. However, the complicated outflows appear to have distinct chemistries generated by shocks and dust destruction. Various RSG envelopes have been probed for their molecular content, mostly by radio and millimeter observations; however, VY Canis Majoris (VY CMa) and NML Cygni (NML Cyg) display the highest chemical complexity, and also the most complicated envelope structure. Thus far, over 29 different molecules have been identified in the envelopes of RSGs. Some molecules are common for circumstellar gas, including CO, SiO, HCN and H$_2$O, which have abundances of $\sim$$10^{-6}$--$10^{-4}$, relative to H$_2$. More exotic oxides have additionally been discovered, such as AlO, AlOH, PO, TiO$_2$, and VO, with abundances of  $\sim$$10^{-9}$--$10^{-7}$. RSG shells support intricate maser emission in OH, H$_2$O and SiO, as well. Studies of isotope ratios in molecules suggest dredge-up at least into the H-burning shell, but further exploration is needed. 
}


\begin{document}
\section{Introduction: Circumstellar Chemistry Beyond the Asymptotic \mbox{Giant Branch}}
\label{sec:intro}
Circumstellar envelopes of evolved stars are among the most remarkable sources of interstellar molecules, producing unique chemical compounds not found anywhere else in the universe \cite{2018A&ARv..26....1H, 2006PNAS..10312274Z}. Their unusual and varied chemical content is in part due to their distinct geometry and physical properties, which are unlike those of molecular clouds. The envelopes are created by mass loss from the stellar photosphere and are characterized by sharp gradients in both temperature and density as a function of radial distance from the star \cite{2001A&A...368..969S, 1988ApJ...326..832K}. The mass loss arises from instabilities in the stellar interior when hydrogen is exhausted in the core, causing it to contract and ignite a surrounding H-burning shell \cite{2005ARA&A..43..435H}. The excess energy generated results in the expansion of the outer atmosphere, which is now highly convective, creating a red giant star. Hot, gaseous matter flows from the star and cools, generating a circumstellar envelope. Within a few stellar radii, dust grains and molecules begin to form in a distinct chemical mixture \cite{1996ARA&A..34..241G}. Once dust forms, radiation pressure on the grains enhances mass loss. Core collapse continues, and, finally, the helium in the stellar center ignites, transforming the central composition to carbon and oxygen \cite{2009ApJ...690..837M}. For lower mass stars (1--8 M$_{\odot}$ ), the red giant phase evolves into the Asymptotic Giant Branch (AGB), in which a He-exhausted carbon and oxygen core is now surrounded by He- and H-burning shells. Mass loss will accelerate on the AGB, creating a substantial, molecule and dust-rich, mostly spherical envelope, such as the famous object {IRC+10216} 
 \cite{2010ApJS..190..348T}. 

The AGB also plays an important role in the proliferation of carbon. When the thermal-pulsing (TP) phase is entered off the early AGB (E-AGB), the convective envelope dips into the He-burning shell and transports $^{12}$C to the stellar surface \cite{2009ApJ...690..837M} in a process called “third dredge-up” \cite{2005ARA&A..43..435H}. This activity flips the C/O ratio from about $\sim$0.5 to $\sim$1.5 in favor of \mbox{carbon \cite{2009ApJ...690..837M}.} Note that circumstellar shells on the TP-AGB are one of the few environments in the interstellar medium where C $>$ O. Here “interstellar” roughly defines material that has left the stellar photosphere and thus is becoming interstellar.

More massive stars (9--40 M$_{\odot}$ ) undergo a different evolutionary path. While on the main sequence, H-burning occurs predominantly through the CNO cycle \cite{2017ars..book.....L}. When hydrogen is depleted in the core, hydrogen ignites in a surrounding shell. The star undergoes core contraction with an expanding envelope, as with lower mass objects. Helium then begins to burn in the core, and the star enters the RSG stage. Weak s-process elements are also produced. Furthermore, the convective envelope will deepen into the interior layers and mix products of nucleosynthesis to the surface.  Eventually, RSGs are sufficiently massive to ignite the carbon accumulated in their cores and progress beyond C-fusion. Even more massive stars follow other evolutionary tracks.  RSGs are typically cool \mbox{(Teff$\sim$3400--4000 K)}  and large  (r$\sim$300--1500 R$_{\odot}$) with high luminosities (log L/L$_{\odot}$)$\sim$4.5--5.71. The envelopes support mass loss rates between 10$^{-6}$--10$^{-4}$ M$_{\odot}$ per year.  RSGs on the higher end of the mass loss and luminosity range are termed Red Hypergiants and Yellow Hypergiants (YHGs), with the latter objects thought to be evolving to higher temperatures \cite{2017ars..book.....L}.

 It has been thought for decades that the majority of massive stars evolve into Type II-P or Type II-L supernovae on the RSG track \cite{2022AJ....163..103H}. However, 20 years of observations have suggested a lack of RSGs above 18 M$_{\odot}$ among SNe progenitors \cite{2015PASA...32...16S, 2022MNRAS.515..897R}, although this point has been debated \cite{2025ApJ...979..117B}. It is thought that there are actually different outcomes for these stars; they could collapse directly to black holes, or evolve back to warmer temperatures before a terminal explosion \cite{2018ApJ...860...93S}.
 
\begin{figure}[H]
  \begin{center}
\includegraphics[width=0.8\textwidth]{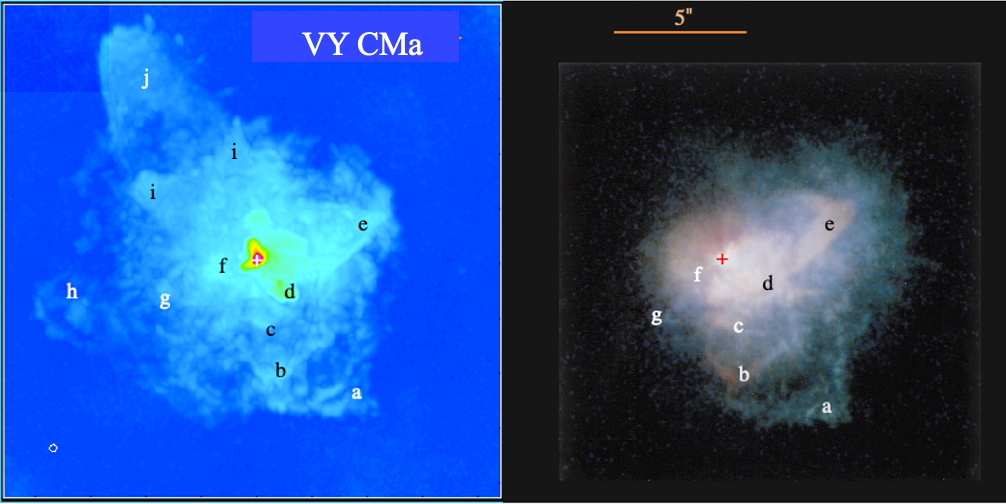}
\end{center}
  \caption{{ALMA} 
 map of $^{12}$CO ($J= 2\rightarrow1$) ({{\textbf{left}}
}) and composite color HST image ({{\textbf{right}}}), of
VY CMa, displayed on the same scale \cite{2023ApJ...954L...1S}. Star position is given by a cross (white:
ALMA; red: HST). The ALMA synthesized beam is shown in the lower left corner; ALMA color scale is (light blue to pink) 0.3-0.7 Jy/beam. Individual features are labeled a--j; some appear in both images but
others that are unique to CO: (a) Arc 1 (b) Arc 2 (c) S Arc (d) SW Clump (e) NW Arc (f) S Knot
(g) SE Loop (h) E Bubble (i) NE Extension (j) NE Arc.
    \label{fig:Mols1}}
\end{figure}

Mass loss from RSGs is impressive and is more complex than the relatively simple, mostly spherical outflows that characterize the AGB. RSGs, in general, appear to undergo sporadic, highly directional mass ejection events occurring over the course of several hundred years that create irregular envelopes consisting of extensive \mbox{{sub-structures}}  
\cite{2007AJ....133.2716H,2021AJ....161...98H,2023A&A...669A..56Q}.
 The famous 
supergiant VY Canis Majoris (VY CMa) is an excellent example of this phenomenon, with an envelope consisting of multiple clumps, knots and arcs that extend \mbox{$>$900 $R_{\star}$} from the central star \cite{2023ApJ...954L...1S, 2009ApJ...695.1604Z}. Very recent maps of CO obtained from the Atacama Large Millimeter Array (ALMA) show an extremely complex structure  well beyond the famous Hubble Space Telescope (HST) image, as shown in Figure~\ref{fig:Mols1} \cite{2023ApJ...954L...1S}. The discrete mass loss phenomenon is thought to result from massive convection cells, perhaps powered by energy stored in twisted magnetic fields \cite{2017ars..book.....L}.

For many years, it was assumed that the envelopes of AGB stars were unique in their rich molecular content, in particular, those with C $>$ O \cite{2010A&A...516A..68K, 2009ApJ...695.1604Z}.  This conclusion was drawn from the numerous molecular studies of these objects at the millimeter and sub-mm wavelengths (e.g., \cite{2018A&ARv..26....1H, 2006PNAS..10312274Z, 2010ApJS..190..348T}).  The envelopes of RGSs were not a major focus, although many such objects were included in surveys of a wide range of circumstellar objects. These works included targeted studies of circumstellar SO and SO$_{2}$ \cite{1993A&A...267..490O}, CN  \cite{1997A&A...319..235B}, and HCN, HNC, SiS, SiO, HC$_{3}$N, and CS \cite{1994A&A...285..247B}. The focus on the shells of C-stars is perhaps understandable. The carbon enrichment from the third dredge-up creates a unique “super-organic” chemistry that leads to many complex C-bearing molecules, including a whole variety of long acetylenic chains not commonly found in molecular clouds \cite{2014ApJ...795L...1A}. Carbon as an element is far more versatile than oxygen in chemical reactions, and the molecular inventory found in C-stars such as IRC+10216 has certainly illustrated this property \cite{2010ApJS..190..348T, 2010A&A...516A..68K, 2009ApJ...695.1604Z}.

\begin{figure}[H]
 \begin{center}
  \includegraphics[width=0.8\textwidth]{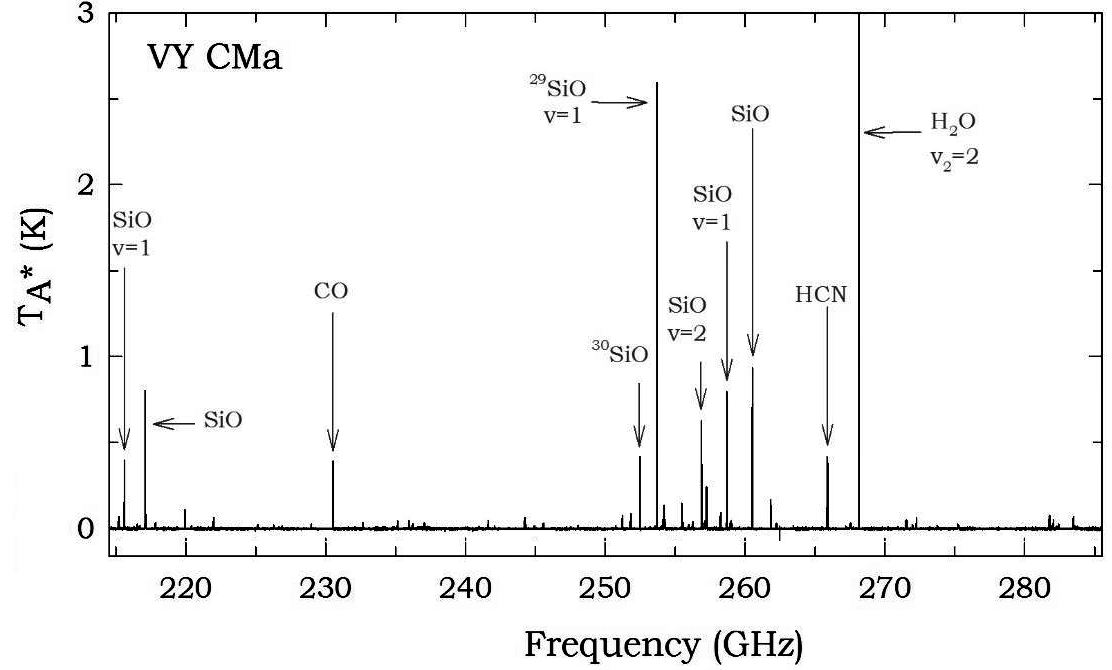}
  \end{center}
  \caption{{The} 
 ARO SMT 1 mm survey of the envelope of Supergiant VY CMa, showing
the stronger features among the 128 individual spectral lines detected \cite{2010ApJS..190..348T}.
    \label{fig:Mols2}}
\end{figure}

This viewpoint changed dramatically over the past decade with new targeted spectroscopic studies of RSGs. Perhaps the most notable is the sensitive spectral-line survey carried out for VY CMa in {2010} 
\cite{2010ApJS..190..348T, 2010ApJ...720L.102T}---the first such survey focused on the chemical characterization of an O-rich circumstellar shell at millimeter wavelengths. Conducted with the Submillimeter Telescope (SMT) of the Arizona Radio Observatory (ARO) with new 1 mm sideband-separating mixers developed for ALMA, this survey showed that the envelope of VY CMa has a complex chemical environment. Over the frequency range 214.5--285.5 GHz of the survey, 128 emission features were detected, arising from 18 different molecules  (see Figure~\ref{fig:Mols2}), with the overall spectrum dominated by SO$_{2}$ and SiS.  In a comparative survey of IRC+10216, 720 lines were observed, assigned to 32 different species in contrast. The observations, however, were conducted with a 30 arcsec beam, favoring IRC+10216 over VY CMa at their respective distances (0.15 kpc vs. 1.2 kpc). Nonetheless, the survey of  VY CMa resulted in the detection of several new interstellar oxide molecules,  including \mbox{PO \cite{2007ApJ...666L..29T},} AlO \cite{2009ApJ...694L..59T}, and AlOH \cite{2010ApJ...712L..93T}. The study also identified the presence of uncommon species such as NaCl and PN in this object. The line profiles observed varied among the detected molecules as well; some spectra had multiple velocity components that appeared to be associated with the highly-directional mass loss events traced by arcs and clumps in HST images \cite{2007Natur.447.1094Z}.   Detections of other new species, such as TiO, TiO$_2$ \cite{2013A&A...551A.113K}, and VO \mbox{followed \cite{2019ApJ...874L..26H}.} Note that oxides like TiO and VO were reported previously  in the optical spectra of stellar photospheres \cite{1971ApJ...169..195W}, but these are the first recognized circumstellar detections.

Other pioneering works on the molecular content of RSGs include \mbox{Teyssier et al. \cite{2012A&A...545A..99T}}. These authors used the HIFI instrument on the \textit{{Herschel Space Observatory}} 
to measure the sub-mm spectra (560--700 GHz; 1100--1201 GHz; 1750--1840 GHz) of the envelopes of four RSG/YHG stars: NML Cyg, Betelgeuse, IRC+10420 and AFGL 2343. They primarily detected high-energy lines of CO, H$_{2}$O, and OH in these objects, with the remarkable identification of NH$_{3}$ towards NML Cyg and IRC+10420. Overall, the high excitation lines were less prominent in the YHGs (IRC+10420, AFGL 2343), suggesting that these objects have hollow, detached envelopes. The sample spectra of Betelgeuse are shown in Figure \ref{fig:Figure2.5}. This study was soon followed by an interferometric 0.8 mm survey of VY CMa, carried out with the Submillimeter Array (SMA) \cite{2013ApJS..209...38K}. In the survey, 223 features in the range 279--355 GHz were identified, arising from 19 molecules, including some like AlCl and TiO, not observed at the SMT. The molecules with the highest number of lines were SO$_{2}$ (53), TiO$_{2}$ ($\geq$33), NaCl (33), SiO (29), and SO (26). 

These works changed the chemical perception of supergiant stars. It was recognized that they could be good molecular “hunting grounds.” As a consequence, more in-depth studies of the molecular content of RSG envelopes were conducted, both in the frequency and spatial domains \cite{2016A&A...592A..51Q, 2024ApJ...971L..43R, 2024A&A...681A..50W}.

\section{The Molecular Inventory of the Envelopes of RSGs}
\label{sec:mol_inventory}
\begin{figure}[H]
 \begin{center}
  \includegraphics[width=0.25\textwidth]{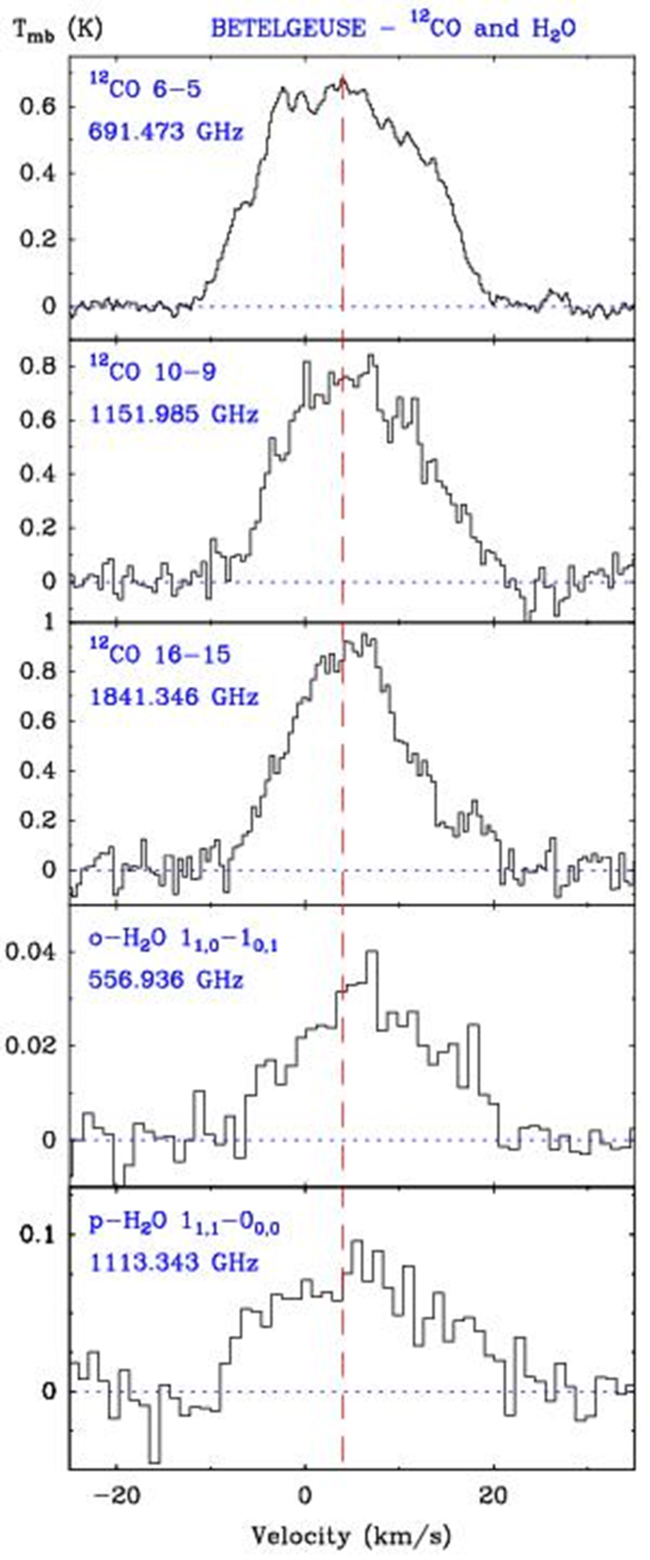}
  \end{center}
     \caption{{Spectra} 
       of high energy transitions of CO and H$_2$O observed towards {Betelgeuese} 
       using \textit{{Herschel}} \cite{2012A&A...545A..99T}.}
    \label{fig:Figure2.5}  
\end{figure}

Currently, several existing spectral-line surveys of RGSs provide a basis for defining the chemical content of these objects.  A sample of representative RSGs and their basic properties is given in Table~\ref{tab:stars}.  For several of these objects, extensive molecular surveys have been conducted at mm wavelengths.  In addition to VY CMa \cite{2010ApJS..190..348T}, NML Cyg, AH Sco, KW Sgr, VX Sgr, S Per and the YHG IRC+10420 have been the subjects of millimeter surveys. In the case of NML Cyg, an identical frequency range was observed with the ARO SMT at 1 mm as  for VY CMa \cite{2022AJ....164..230S, 2021ApJ...920L..38S}. A survey was also conducted for IRC+10420 at both 3 mm (83--117 GHz) and 1 mm (199--277 GHz) employing the Institute de Radioastronomie Millimetrique (IRAM) 30-m telescope \cite{2016A&A...592A..51Q}, and for AH Sco, KW Sgr, and VX Sgr using ALMA in the range $\sim$213--270 GHz, as part of the larger ATOMIUM project to study O-rich circumstellar shells \cite{2024A&A...681A..50W}. The later work focused on the spectral line data, presenting some general spatial information and images of SO and SO$_2$. In the case of Betelgeuse, molecular observations were conducted at 0.8 mm \cite{2018A&A...609A..67K}, as well as those with \textit{{Herschel}}, as mentioned (also see Figure  \ref{fig:Figure2.5}). Limited molecular observations also exist for S Per \cite{1993ASPC...35..199K, 2014apn6.confE..72Q}. SiO and H$_2$O appear to be ubiquitous, in the IR stellar atmosphere or maser detections as well as existing mm-wave surveys. Nevertheless, these works provide an important window on the molecular content of envelopes of RSGs. 
Table~\ref{tab:mols} summarizes the current set of molecules identified in circumstellar gas of RSGs, a total of 29 different chemical compounds. The listed species are secure detections, and all except VO were identified on the basis of their pure rotational spectrum measured at mm/sub-mm wavelengths. Many of the molecules found are common species in C- and O-rich AGB envelopes, including CO, CS, SiO, SiS, and HCN \cite{2010ApJS..190..348T, 2017A&A...597A..25V}.

\begin{figure}[H]
\begin{center}
  \includegraphics[width=\textwidth]
  {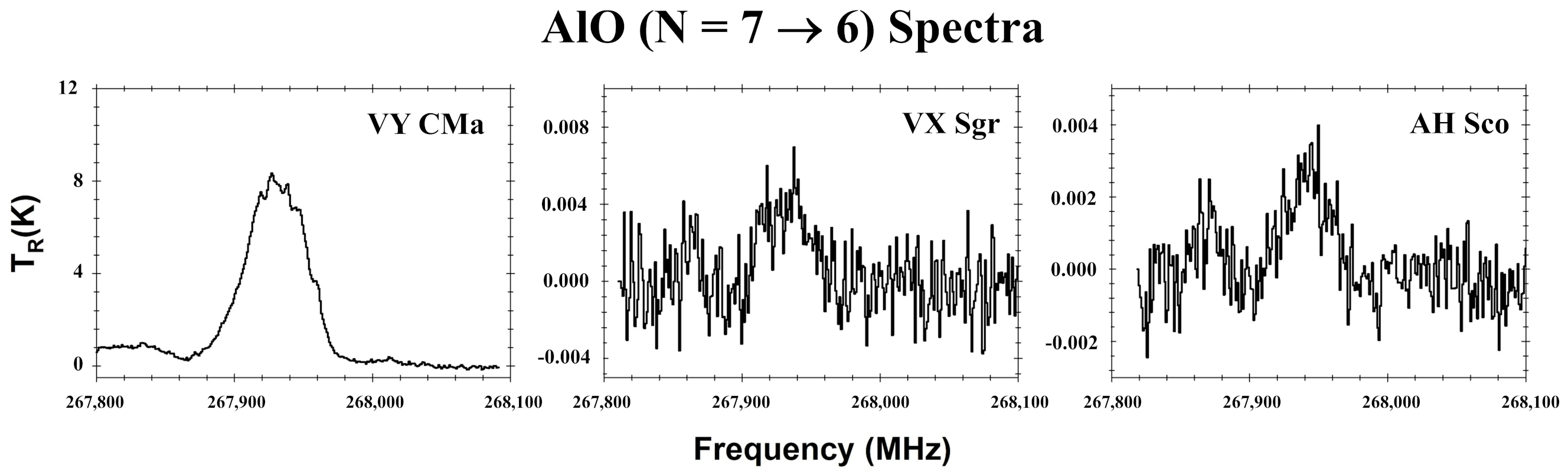}
  \end{center}
   \vspace{-10pt}
   \caption{{Circumstellar} 
 spectra of the $N = 7\rightarrow6$ transition of AlO measured
towards the envelopes of three RSG stars, using ALMA: VY CMa, VX Sgr, and
AH Sco. The broad profiles are in part due to the fine and hyperfine structure in
AlO. The line is strongest in VY CMa \cite{2024A&A...681A..50W, Ravi2025b}.
    \label{fig:Mols3}}
\end{figure}

\begin{figure}[H]
\begin{center}
  \includegraphics[scale=0.8]{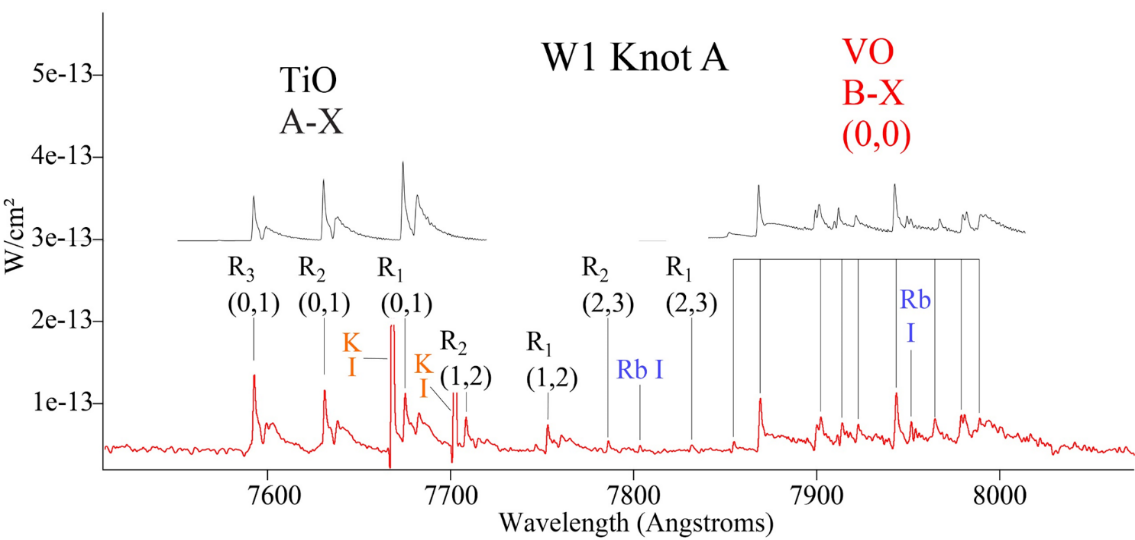}
  \end{center}
  \caption{{HST}
 spectra of the circumstellar gas about 50 $R{_\star}$ from the star near 7800 \AA\,  (W1 Knot A), showing the A-X and B-X electronic transitions of TiO (left) and VO
(right), identifying VO as a new interstellar molecule \cite{2019ApJ...874L..26H}. The numbers in parentheses
are the respective vibrational quantum numbers of the respective states, and R$_1$, R$_2$, etc.,
are the spin-orbit ladder components. Simulated spectra are shown above the data. K{\sc i}
atomic lines in the TiO data have been truncated.
    \label{fig:Mols4}}
\end{figure}

\begin{table}[H]
\caption{{Representative} 
 Red Supergiant Stars. \label{tab:stars}}

\small
\begin{adjustwidth}{-\extralength}{0cm}
\begin{tabularx}{\fulllength}{LLLLLLL}
\toprule
\textbf{Object}&\textbf{Mass}\linebreak\textbf{(M}\boldmath{$_{\odot}$}\textbf{)} &{\textbf{Mass Loss Rate}}\linebreak \textbf{(M}\boldmath{$_{\odot}$} \textbf{yr}\boldmath{$^{-1}$}\textbf{)}&\textbf{Envelope Structure} &\textbf{Molecular Content}&\textbf{Comments}&\textbf{Spectra} \\

\midrule
VY CMa \s{a}    &   40       & $6\times10^{-4}$       & Asymmetric; highly complex  & Rich in metal oxides, halides, PO, PN& Red Hypergiant&Multiple blue-, red-shifted features\\
\midrule
NML Cyg \s{b}   &    25      &$1\times10^{-4}$        &Asymmetric                   & Rich in metal oxides, halides, PO, PN& Red Hypergiant&Multiple blue-, red-shifted features\\
\midrule
Betelgeuse \s{c}&12--25      &$2\times10^{-6}$        &Recent flares                & CO, HCN, OH, H$_2$O                           &RSG          &Mostly symmetric \s{h}\\
\midrule
IRC+10420 \s{d} &20          &$5\times10^{-4}$ \s{g}   &Roughly spherical            &N-rich, no “metal” species          &Yellow Hypergiant& Symmetric \\
\midrule
AH Sco \s{e}    &     20      &$1\times10^{-5}$        &Spherical                    &Metal oxides, AlF, PO&    RSG              &  Broad and narrow features \\
\midrule
VX Sgr \s{e}   &      12     &$6\times10^{-5}$        &Spherical, shell-like        &Metal oxides, AlF, PO&    RSG              & Symmetric, broad and {narrow} \\ 
\midrule
KW Sgr \s{e}   &      20     &$3\times10^{-6}$        &                  --         &CO, HCN, SO          &    RSG              & Weak\\
\midrule
S Per \s{f}       & 30  & $2.4\times10^{-5}$   &Spherical, some axial-symmetry& CO, H$_2$O, OH \s{i}         & RSG                & -- \\
\bottomrule
\end{tabularx}
\end{adjustwidth}
\noindent{\footnotesize{\s{a} \cite{2023ApJ...954L...1S, 2016AJ....151...51S, 2010ApJ...720L.102T};
{\s{b}} 
 \cite{2022AJ....164..230S, 2021ApJ...920L..38S};      
\s{c} \cite{2018A&A...609A..67K};                                               
\s{d} \cite{2016A&A...592A..51Q};                                  
\s{e} \cite{2024A&A...681A..50W};                                          
\s{f} \cite{1976A&A....50..153G, 2018AJ....155..212G, 1993ASPC...35..199K}; 
\s{g} Previous mass loss rate $\sim$10$^{-3}$ M$_{\odot}$ yr$^{-1}$}; 
\mbox{\s{h} {On scales $\lesssim5^{''}$}; }
\s{i} Northern source, no published wide-range survey.}
\end{table}
 Despite the fact that RSG envelopes are oxygen-rich, carbon chemistry is still active, as represented by the presence of HCN, CN, HNC and HCO$^+$ \cite{2010ApJS..190..348T, 2009ApJ...695.1604Z, 2013ApJS..209...38K, 2022AJ....164..230S}. Furthermore, sulfur-bearing molecules, including SO$_2$, SO and SiS, appear to dominate the spectra of these \mbox{objects \cite{2010ApJ...720L.102T, 2013ApJS..209...38K, 2024A&A...681A..50W, 2022AJ....164..230S}.} Perhaps  most striking is that $\sim$20\% of the species  listed in Table~\ref{tab:mols} are their first identifications in the ISM;  in particular, the refractory oxides AlO, PO, AlOH, TiO, TiO$_2$, and VO were all first discovered towards VY CMa \cite{2007ApJ...666L..29T, 2009ApJ...694L..59T, 2010ApJ...712L..93T, 2013A&A...551A.113K, 2019ApJ...874L..26H}.  With the exception  of VO, these species have subsequently been found in AGB envelopes, such as IK Tau and Mira \cite{2020MNRAS.494.1323D, 2018A&A...620A..75K}, as well as in other RSGs. Note that Wallerstein observed the electronic transitions of VO in the photosphere of VY CMa \cite{1971ApJ...169..195W}. AlO spectra, measured in VY CMa, VX Sgr, and AH Sco  with ALMA, are shown in Figure~\ref{fig:Mols3}. VO currently remains  the only species found solely in VY CMa. Unlike all the other molecules, VO was observed via its B-X electronic transition in gas about 50 $R_{\star}$ from the star \cite{2019ApJ...874L..26H}; see Figure~\ref{fig:Mols4}.  The special geometry needed for stellar excitation of VO may not occur in other RSG envelopes. It is interesting to note some of the “non-detections,” including certain metal oxides (FeO and MgO) and refractory hydrides (SiH, PH) \cite{2020ApJ...901...22S}.

\begin{table}[H]
  \caption{{Molecules} 
 Detected in the Envelopes of RSGs \s{a}.
  \label{tab:mols}}
\begin{tabularx}{\textwidth}{LLLLLLLLLL}
\toprule
  H$_2$&
HCN&
SiO&
PN&
AlO&
CO&
HCO$^+$&
SiS&
PO&
VO\\
CS&
HNC&
SO$_2$&
NaCl&
AlOH&
CN&
H$_2$O&
SO&
KCl&
TiO\\
NO&
\mbox{N$_2$H$^+$ \s{b}}&
NS&
AlCl&
TiO$_2$&
OH&
NH$_3$&
H$_2$S&
AlF\\
\bottomrule
\end{tabularx}

\noindent{\footnotesize{\s{a} All millimeter-wave detections except VO (see text); \s{b} IRC+10420 only}.}
\end{table}
\vspace{-6pt}

\begin{figure}[H]
 \begin{center}
  \includegraphics[width=0.85\textwidth]{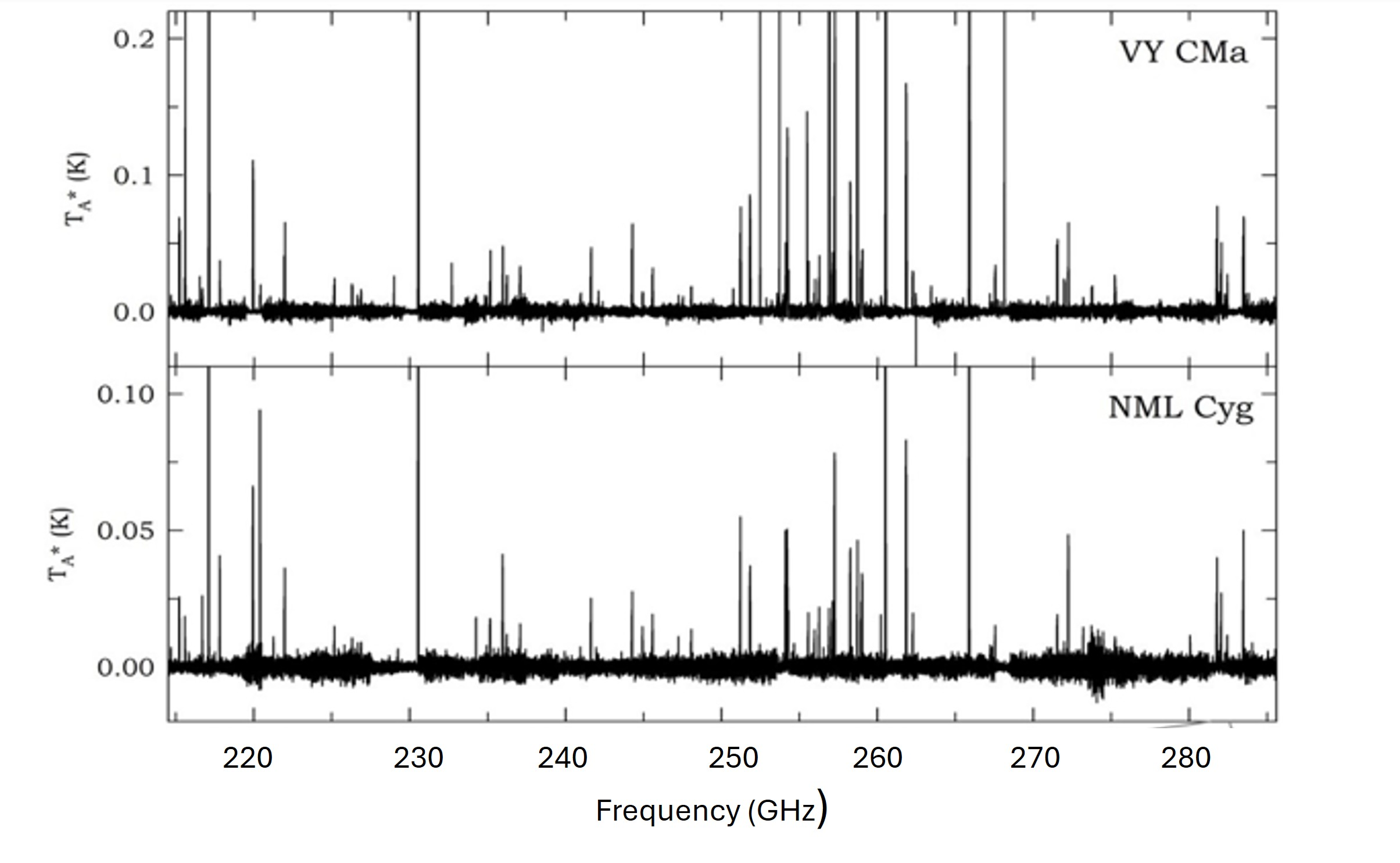}
  \end{center}
  \vspace{-20pt}
  \caption{{Composite} 
 spectrum of the ARO SMT 1 mm survey of VY CMa (\textbf{upper panel}; also
Figure~\ref{fig:Mols1}), and an identical survey spectrum of NML Cyg (\textbf{lower panel}), covering 215--285
GHz. The spectra look almost identical \cite{2010ApJS..190..348T, 2022AJ....164..230S, 2021ApJ...920L..38S}.
    \label{fig:Mols5}}
\end{figure}

Not all the molecules in Table~\ref{tab:mols}, however, are found in every RSG envelope. Almost all the species listed have been identified in the circumstellar gas of VY CMa, with the exception of N$_2$H$^+$. VY CMa is in fact the most chemically diverse of all RSGs and is the only one thus far known to contain VO, AlCl and TiO$_2$ \cite{2013A&A...551A.113K, 2019ApJ...874L..26H}. The current non-detection of N$_2$H$^+$ in VY CMa may be an observational selection effect, because the frequency is at the high end of the ALMA Band 6. NML Cyg has almost an identical set of molecules as VY CMa, as made evident in a 1 mm spectral survey \cite{2022AJ....164..230S}, conducted over the same frequency range as the original one for VY CMa. In fact, as shown in Figure~\ref{fig:Mols5}, the \mbox{1 mm} composite spectra of NML Cyg and VY CMa, truncating the strongest lines, look remarkably similar over the frequency range 214.5--285.5 GHz.   The only molecule not found in NML Cyg in the inventory of VY CMa is AlOH. This source, however, is more distant than VY CMa (1.6 kpc vs. 1.2 kpc), reducing the sensitivity for single-dish observations. It is likely that with further integration,  AlOH will be identified in NML Cyg. The molecule has been found in other RSG and AGB envelopes, such as those of VX Sgr and AH Sco \cite{2024A&A...681A..50W}.

\vspace{-6pt}

AH Sco and VX Sgr have been found to contain the majority of molecules in \mbox{Table~\ref{tab:mols}  \cite{2024A&A...681A..50W}.} They appear to be rich in aluminum compounds, including AlO, AlOH and AlF, and also contain TiO. However, the alkali metal halides NaCl and KCl are not present in either source. In contrast, both Na$^{35}$Cl and Na$^{37}$Cl are observed in both VY CMa and NML Cyg, including the $v$ = 1 and 2 vibrationally-excited states of the main \mbox{isotopologues \cite{2010ApJS..190..348T, 2013ApJS..209...38K, 2022AJ....164..230S, 2021ApJ...920L..38S}.} In VY CMa, both NaCl and KCl trace gas not only near the star, where they are produced by LTE chemistry \cite{1973A&A....23..411T}, but they are also observed in the SW {Knots} 
\cite{2025AJ....169..230H}, which comprise the material ejected from the star about 200 years ago. The absence of both these species in AH Sco and VX Sgr is puzzling.

The YHG IRC+10420 has a distinct subset of molecules from Table~\ref{tab:mols} \cite{2016A&A...592A..51Q}. First, it is the only supergiant thus far with the ion N$_2$H$^+$, and has relatively strong emission from NO. There is also a tentative detection of CH$_3$OH in this source \cite{2016A&A...592A..51Q}. On the other hand, unlike RSGs, IRC+10420 lacks metal-containing molecules---metals in the chemist’s sense. No aluminum, titanium, sodium or potassium-bearing species have been thus far reported in this source.  The refractory molecule PN has been identified, but not PO. In fact, over half the molecules observed in IRC+10420 are nitrogen-bearing, including\mbox{ NH$_3$ \cite{1995ApJ...448..416M}}. {Quintana-Lacaci et al.} \cite{2016A&A...592A..51Q} 
concluded that the envelope exhibits an N-based chemistry, with excess nitrogen generated by nucleosynthesis within the star. Excess nitrogen can be created by dredge-up from the H-burning shell, which would coincide with elevated $^{13}$C \cite{1973ApJ...184..493A}. The $^{12}$C/$^{13}$C ratio found in the envelope is quite low ($\sim$4--7) and is further evidence
\noindent for such mixing. Perhaps the metals are contained in species such as AlN and TiN. Additional line searches would be informative, although accurate rest frequencies for metal nitrides are generally not available.

  \begin{figure}[H]
\begin{center}
  \includegraphics[width=0.6\textwidth]{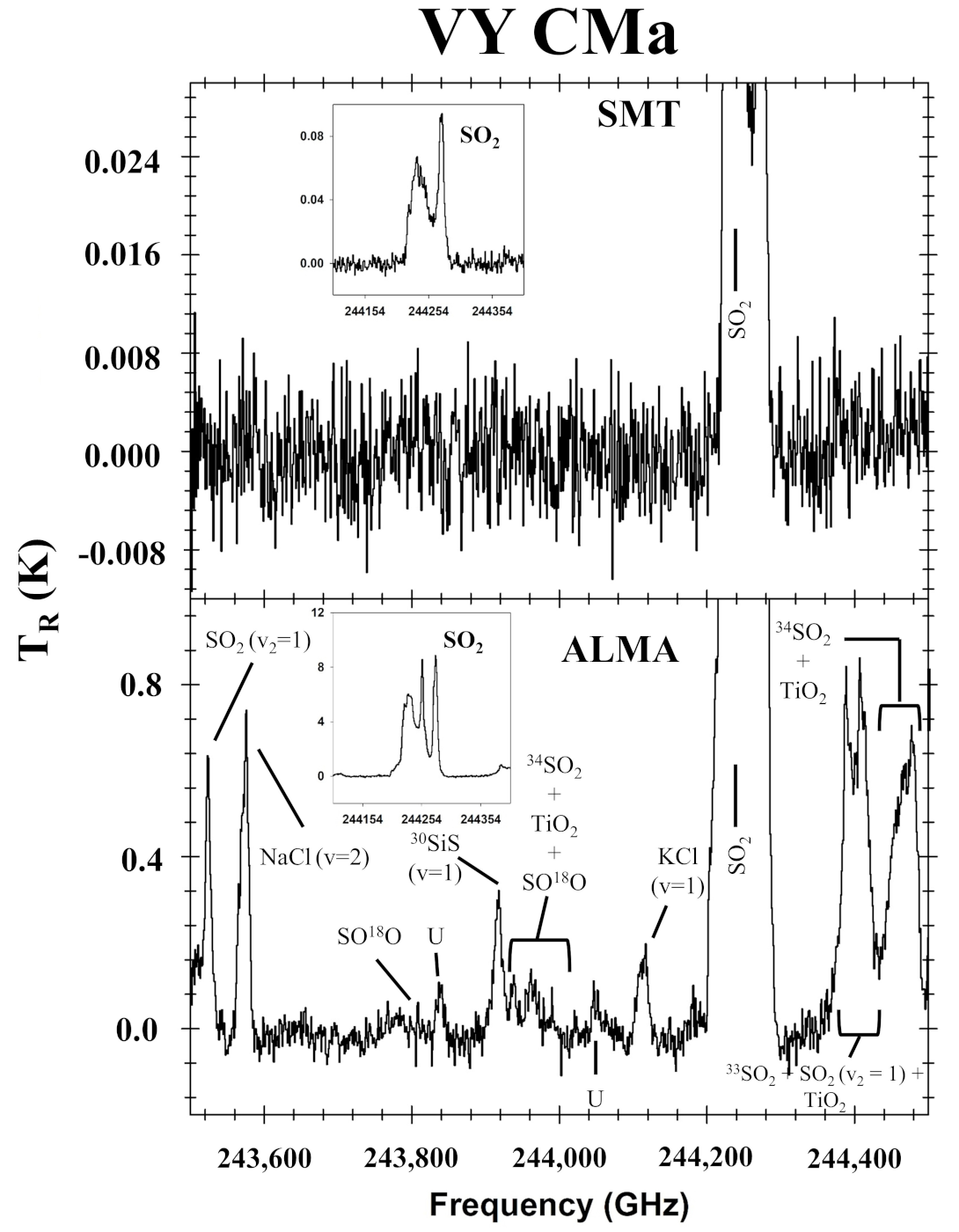}
  \end{center}
  \vspace{-10pt}
  \caption{{Comparison} 
 of spectra of VY CMa from the SMT 1 mm survey
    ({{\textbf{upper}}}) and from ALMA ({{\textbf{lower}}}), near 244 GHz \cite{2010ApJS..190..348T, Ravi2025a}. 
    The truncated line is SO$_2$, shown in the insets. Note the sharp blue and red-shifted components of SO$_2$.
    \label{fig:Mols6}}
\end{figure}
 
The envelopes of Betelgeuse and KW Sgr appear to be less complex chemically than the other stars. They contain CO, H$_2$O, SiO, HCN, and presumably H$_2$, and SO has been observed in KW Sgr \cite{2024A&A...681A..50W, 2018A&A...609A..67K}. However, common species such as CS and HCO$^+$ have not been yet detected in these objects. The lack of molecules may arise from the lower mass loss rates, as will be further discussed.

Many of these envelopes also contain unidentified lines (U-lines). In the ARO 1 mm surveys \cite{2010ApJS..190..348T, 2021ApJ...920L..38S}, 14 U-lines were found towards the envelope of VY CMa and 4 in that of NML Cyg. Because of the large ARO SMT beam ($\sim$30 arcsec), these studies are not as sensitive as those from ALMA. More unidentified lines are likely to be found in these shells. A glimpse at the actual spectral line density is shown in Figure~\ref{fig:Mols6}. Here, a $\sim$1 GHz wide spectrum is presented at 244 GHz, comparing data from ALMA \cite{Ravi2025a} 
and the ARO SMT survey \cite{2010ApJS..190..348T}, lower and upper plots, respectively. In the SMT spectrum, only the $J_{\mathrm{Ka,Kc}}  = 14_{0,14}\rightarrow13_{1,13}$ line of SO$_2$ is visible, while that of ALMA shows at least 10 additional lines, some of which remain unidentified. The other identified features arise from isotopologues of SO$_2$ and from vibrationally-excited lines of NaCl and KCl. The presence of unknown spectral features suggests a further need for laboratory measurements. 

\section{The Surprising Variation in Spatial Distributions of Molecules }
\label{sec:mol_vary}
Clues to the chemistry of RSGs lie in the spatial relationship of the various molecules. One of the early attempts to examine the relative distributions of the chemical compounds in VY CMa was by Kaminski et al. \cite{2013ApJS..209...38K}, whose pioneering work was conducted with the SMA. These authors found some correlation between molecular emission and the HST dust features, but did not produce high-fidelity maps due to the loss of flux. They estimated as much as 50\% of the emission for certain molecules was resolved out (on scales $\gtrsim4^{''}$). This issue is overcome in new ALMA images of VY CMa, conducted in Band 6 with 0.2$^{''}$--1.5$^{''}$ resolution \cite{2023ApJ...954L...1S, 2024ApJ...971L..43R, Ravi2025a}.
These maps include the addition of single-dish data, and therefore, recover all the flux, presenting a ``complete picture''. Outside of the VY CMa images, little else exists in the current literature for RSGs.  NML Cyg is too far north to be observed with ALMA. The ATOMIUM survey has ALMA data for KW Sgr, VX Sgr and AH Sco in many molecules.  A general comparison of spatial distributions across all 17 envelopes   has been published \cite{2024A&A...681A..50W}, but only maps for SO and SO$_2$ (note neither species was detected in KW Sgr), although all image cubes and other material are available from the ALMA Science Archive. In the case of IRC+10420, molecular images exist only for SiO \cite{2017ApJ...851...65D} and OH \cite{2005MmSAI..76..467R}.

It is unlikely that the envelopes of all RSGs resemble VY CMa in the variable spatial distributions of molecules. For example, images of SO$_2$ in VX Sgr and AH Sco are fairly spherical, with some core/halo structure. Furthermore, the molecular line profiles of many RSGs are fairly symmetric. They do not appear to show evidence of sharp blue- and red-shifted components as observed for VY CMa and NML Cyg. These multiple spectral components in VY CMa and NML Cyg indicate a complex envelope structure (see Figure~\ref{fig:Mols6}; Table~\ref{tab:stars}).
Nevertheless, images of VY CMa provide insight into the extent of the complexity of the molecular envelope, and into the chemical processes that are occurring. 

\begin{figure}[H]
\begin{center}
\includegraphics[width=0.9\textwidth]{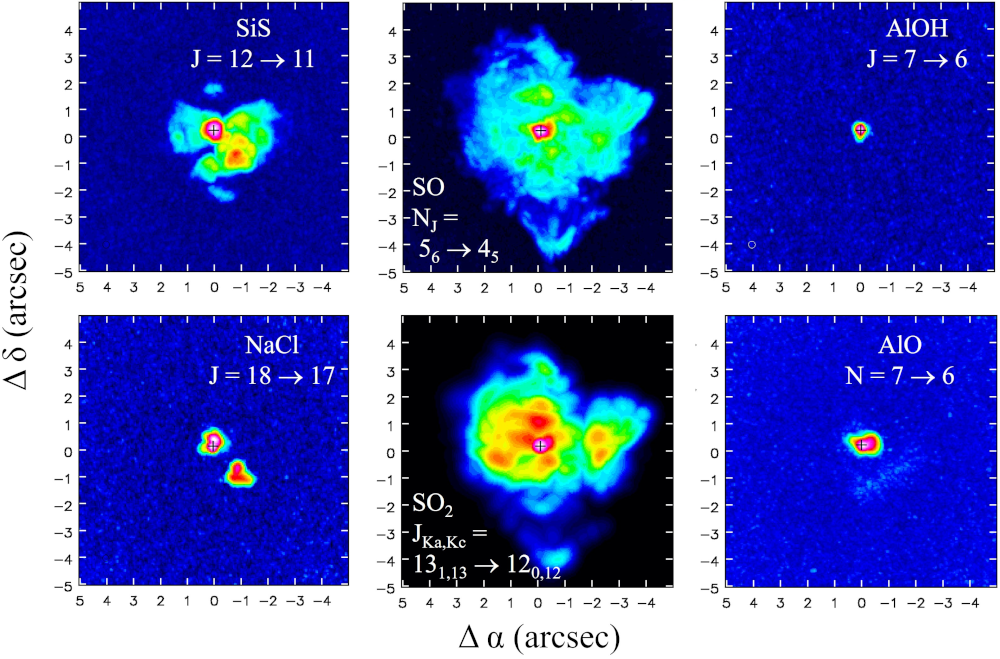}
 \vspace{-15pt}
  \end{center}
  \caption{{Combined} 
 ALMA/SMT images of VY CMa of SiS ($J =12\rightarrow 11$), SO ($N_J =5_6\rightarrow4_5$), AlOH ($J = 7\rightarrow 6$), NaCl ($J =18\rightarrow 17$), SO$_2$ ($J_{\mathrm{Ka,Kc}} = 13_{1,13}\rightarrow 12_{0,12}$), and AlO ($N= 7\rightarrow 6$),
    plotted on the same spatial scale \cite{2024ApJ...971L..43R, Ravi2025b}.
    Crosses indicate the stellar position. Maximum
flux (Jy beam$^{-1}$; in pink) is 0.6 for SO and SiS, 0.45 for SO$_2$, 0.14 for NaCl, 0.12 for AlO, and 0.024 for AlOH. AlO and AlOH are centered on the star, NaCl and SiS also highlight the SW
Clumps, while SO and SO$_2$ trace more extended structures.
\label{fig:Mols7}}
\end{figure}

\begin{figure}[H]
\begin{center}
\includegraphics[width=0.9\textwidth]{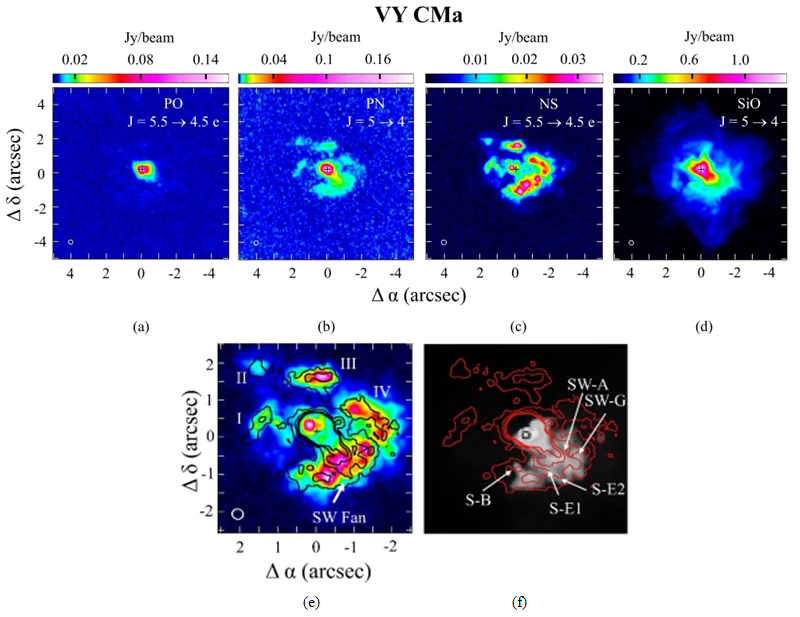}
  \end{center}
  \vspace{-15pt}
  \caption{{\textbf{Upper (a, b, c, d):}} 
 Combined ALMA/SMT images of PO ($J=5.5 \rightarrow 4.5e$), PN ($J=5\rightarrow4$), NS ($J=5.5\rightarrow4.5e$), and
SiO ($J=5\rightarrow4$), left to right. Synthesized beam sizes are shown at lower left as white circles; crosses indicate the stellar
position. Flux scale (Jy beam$^{-1}$) is indicated on the figure. PO is clearly centered on the star,
while PN extends to the SW with a close correspondence to NS and partly to SiO. {\textbf{Lower}} (\textbf{{e}},\textbf{{f}}):  Emission contours of PN superimposed over ({\textbf{left}}) image of NS from above (in black, ({\textbf{e}})) and ({\textbf{right}}) the HST image (in red, ({\textbf{f}})). The SW Fan, Cloudlets I, II, III, and IV and certain HST dust clumps are indicated. PN emission coincides with NS and some of the dust clumps, including South Knots B, E1 and E2 and SW Knots A and G \cite{2024ApJ...971L..43R}.
    \label{fig:Mols8}}

\end{figure}

A representative sample of ALMA images of various molecules in VY CMa is shown in Figures~\ref{fig:Mols7} and~\ref{fig:Mols8}. Note that the map in CO is given in Figure~\ref{fig:Mols1}. Figure~\ref{fig:Mols7} displays ALMA images of SiS ($J =12\rightarrow11$), SO ($N_{\mathrm J} =5_6\rightarrow4_5$), AlOH ($J = 7\rightarrow6$), NaCl ($J =18\rightarrow17$), SO$_2$ ($J_{\mathrm{Ka,Kc}} =  13_{1,13}\rightarrow12_{0,12)}$, and AlO ($N= 7\rightarrow6$),  plotted on the same spatial \mbox{scale \cite{2024ApJ...971L..43R, Ravi2025b}.}
Note that an earlier ALMA image of NaCl had been published, but with less \mbox{sensitivity \cite{2016A&A...592A..76D}}. In Figure~\ref{fig:Mols8}, ALMA images of PN, PO, NS and SiO are shown, also on the same spatial scale. The intensity scale (in Jy beam$^{-1}$) is indicated on the figures or in the figure captions. Note that these maps all contain the single-dish addition.
 \vspace{10pt}

The most remarkable result from these images is the large differences in spatial distributions among chemical species. Because the dipole moments and excitation energies of these molecules are very similar, these variations in spatial location are primarily the result of chemistry and, in certain cases, elemental enrichment. AlO and AlOH are only found around the star, within about 60 $R_{\star}$. In sharp contrast, NaCl and SiS are present not only near the star but also in a clump to the southwest, the so-called ``Southwest Clump'', see Figure~\ref{fig:Mols1}, ``d''. The ``SW Clump'' is actually a cluster of clumps \cite{2025AJ....169..230H}, and NaCl has at least two velocity components in this structure corresponding to the knots ``E1/E2`` and the ``S Arc''. The HST dust structures Arc 2 and the NW Arc are traced well by SO and SO$_2$, and are  visible as spectral features in the observed line profiles of these molecules \cite{2013ApJ...778...22A}, see Figure~\ref{fig:Mols6}. Their emission also traces some of the more extended gas seen in CO (Figure~\ref{fig:Mols1}). SiS is more confined than SO and SO$_2$, and is present in the bridge structure called the ``SW Fan'',  and a few other smaller features, resembling the distribution of NS (see Figure~\ref{fig:Mols8}).

In Figure~\ref{fig:Mols8}, the distributions of the two phosphorus-bearing molecules are shown: PO and PN (a, b). The extent of oxide PO closely resembles that of AlO and is centered on the star, extending out to $\sim$60 $R_{\star}$. PN has a far more intricate structure, with the continuous emission SW of the star in a fan-like geometry, the ``SW Fan'' as mentioned (see Figure~\ref{fig:Mols8}e). The fan encompasses part of the SW Clump cluster. PN appears to trace the cluster knots S-B, SW-A, SW-G, S-E1/E2 (see Figure~\ref{fig:Mols8}f) \cite{2024ApJ...971L..43R, 2025AJ....169..230H}. PN also exhibits four other small structures to the north, east and west of the star (``Cloudlets'' I, II, III, IV; see e). These features are even more conspicuous in NS, as the figure shows (see c, e). The data suggest a strong connection between the N-bearing species PN and NS, which will be discussed in Section~\ref{sec:phosphors}. SiO is also observed in the SW Fan and in some of the Cloudlets but has emission  extending over 4$''$ from the stellar position, faintly tracing the Arc 2 and the NW Arc, among other features (see Figure~\ref{fig:Mols1}). SiO is thought to be a tracer of shocks, and may highlight shocked regions \cite{1989ApJ...343..201Z}.

Because these images recover all of the flux, they genuinely illustrate the complexity of the chemical composition of the envelope of VY CMa. Some species are spatially correlated, for example, AlO and PO near the star, or  PN and NS in the SW Fan and the Cloudlets. Other molecules appear to intricately trace the dust features to the south and west, where there is less extinction, such as CO and SO \cite{2023ApJ...954L...1S}. The full impact of these data is\mbox{ yet to be determined.}

\section{The Unusual Chemistry in RSGs}
\label{sec:mol_chem}
\subsection{An Overview of Abundances}
\label{sec:abundances}
The envelopes of RSGs offer a unique and energetic chemical environment where \mbox{C/O $<$ 1 \cite{2009ApJ...690..837M}.} This excess of oxygen has a unique effect on chemistry. The molecular composition favors metal oxides and hydroxides (AlO, AlOH, TiO, TiO$_2$), which are not observed in molecular clouds \cite{2006PNAS..10312274Z}, nor in C-rich AGB stars \cite{2010ApJS..190..348T}. The O/C ratio favors oxygen in RSGs because carbon is not convectively mixed to the surface from the He-burning shell, as mentioned, which occurs on the AGB \cite{2009ApJ...690..837M}.

A summary of molecular abundances, relative to H$_2$ 
is given in Table~\ref{tab:abundances}, which lists those for VY CMa, NML Cyg, and IRC+10420, as compiled from the literature. Uncertainties are estimated at 20--40\%. For several interesting RSGs, the abundances have not yet been published. However, these three stars provide a representative sample. It should be noted that these abundances have not been derived from the same method. The majority of those values for VY CMa and NML Cyg have been determined from a radiative transfer code, ESCAPADE, that considers both spherical and asymmetric outflows, as described \mbox{in \cite{2013ApJ...778...22A}.}  The advantage of this modeling is that the lines profiles are fit to derive both peak abundance and spatial extent. The abundances for IRC+10420 are from a rotational diagram analysis, which assumes LTE and source sizes of 11'' for CO and 3'' for all other \mbox{molecules \cite{2016A&A...592A..51Q}.} Although the latter is not as accurate as a radiative transfer analysis, it nonetheless provides insight into abundances. References for all quoted abundances are given in the table.

As seen in Table~\ref{tab:abundances}, it is clear that most of the C-bearing molecules have about the same abundances across all three stars. CO has $f$$\sim$(1--5) $\times$ $10^{-4}$ while HCN and HNC exhibit $f$$\sim$(1-8) $\times$ $10^{-6}$ and $f$$\sim$ (3--9.7) $\times$ $10^{-8}$, respectively. The abundances for CS and HCO$^+$ are also similar among the stars, with $f$(CS/H$_2$)$\sim$(0.3--2) $\times$ $10^{-7}$ and \mbox{$f$(HCO$^+$/H$_2$)$\sim$ (0.7--4) $\times$ $10^{-8}$. }There is more disagreement for CN; while VY CMa and NML Cyg have $f$$\sim$ (0.6--2) $\times$ $10^{-7}$, the value in IRC+10420 is higher at $1.3 \times 10^{-6}$; the higher value may be due to the proposed nitrogen enrichment in this object \cite{2016A&A...592A..51Q}. There is also a broad range of values for SiS: (0.07--2) $\times$ $10^{-6}$. It is more difficult to quantify the abundance of SiO, as the lines are often non-thermal, but it is greater than $10^{-6}$. The abundances of SO and SO$_2$ are consistent among the three sources, with $f$$\sim$(0.4--1.1) $\times$ $10^{-6}$ and $f$$\sim$(3.4--6.5) $\times$ $10^{-7}$, within a factor of less than 3. On the other hand, those of PN vary by a factor of 20, (0.4--8) $\times$ $10^{-8}$. In general, the most abundant species is CO, followed by H$_2$O, SiO and HCN, as best can be determined from the data. The high abundances of CO, H$_2$O and SiO are perhaps expected, because the stars are oxygen-rich. The prominence of HCN is surprising, as it is not predicted to be prevalent where  C $<$ O. This species has a similar high abundance in oxygen-rich AGB stars (see Table~\ref{tab:abundances}). 
Ziurys et al. \cite{2009ApJ...695.1604Z} suggested that the formation of the molecule may arise from CN generated by shocks, created by the sporadic, energetic mass loss events, or by stellar pulsations \cite{2006A&A...456..549D,1998A&A...330..676}. Even in material with C $<$ O, reactive carbon influences the chemistry.

\begin{table}[H]
  \caption {Typical Molecular Abundances $f$ in the Representative Envelopes of RSGs $^{\text{a}}$.
    \label{tab:abundances}}

\begin{adjustwidth}{-\extralength}{0cm}
 \begin{tabularx}{\fulllength}{ LLLLL}
    \toprule
    \textbf{Molecule}     & \textbf{VY CMa} \boldmath{$^{\text{b}}$}    & \textbf{NML Cyg} \boldmath{$^{\text{c}}$}     & \textbf{IRC+10420} \boldmath{$^{\text{d}}$}     & \textbf{O-Rich AGB} \boldmath{$^{\text{e}}$} \\
    \midrule
CO            & $3.2\times10^{-4}$     & $\sim$$10^{-4}$           & $5.3\times10^{-4}$     & (1--5) $\times$ $10^{-4}$\\
HCN           & (1--8) $\times$ $10^{-6}$ & $2\times10^{-6}$     & $\ge$$1.1\times10^{-6}$     & (0.9--9) $\times$ $10^{-6}$\\
HNC           & $9\times10^{-8}$     & $3\times10^{-8}$     & $9.7\times10^{-8}$     & $-$\\
CN            & $6\times10^{-8}$     & $2\times10^{-7}$     & $1.3\times10^{-6}$     & (0.002--2) $\times$ $10^{-7}$\\
SO$_2$        & $6\times10^{-7}$     & $6.5\times10^{-7}$     & $3.4\times10^{-7}$     & $10^{-6}$\\
SO            & $3.5\times10^{-7}$     & $3.5\times10^{-7}$     & $1.1\times10^{-6}$     & $10^{-6}$\\
H$_2$O     & 2.4 $\times$ $10^{-4}$ $^{\text{f}}$ & $>$$4\times10^{-5}$ $^{\text{g}}$     & $-$     &(1--2)  $\times$  $10^{-4}$ \s{k}\\
HCO$^+$ $^{\text{h}}$  & $3\times10^{-8}$     & $4\times10^{-8}$     & $\ge$$6.7\times10^{-9}$     & (2--10) $\times$ $10^{-8}$\\
N$_2$H$^+$    & $-$     & $-$     & $1.5\times10^{-8}$     & $-$\\
CS            & $2\times10^{-7}$     & $3.0\times10^{-8}$     & $1.2\times10^{-7}$     & (0.07--8) $\times$ $10^{-7}$\\
SiO          & $3\times10^{-5}$ & $\sim$$10^{-6}$     & $\ge$$1.3\times10^{-6}$     & (0.04--2) $\times$ $10^{-5}$\\
SiS           & $2\times10^{-6}$     & $2\times10^{-7}$     & $7\times10^{-8}$     & (0.001-1) $\times$ $10^{-5}$\\
H$_2$S        & $4\times10^{-7}$     & $4.5\times10^{-6}$     & $-$     & (0.04--2.5) $\times$ $10^{-5}$\\
NaCl          & $3.5\times10^{-9}$     & $4.5\times10^{-9}$     & $-$     & (0.05--3) $\times$ $10^{-7}$\\
PN            & $8\times10^{-8}$     & $4\times10^{-9}$     & $3.7\times10^{-8}$     & (1--2) $\times$ $10^{-8}$ $^{\text{i}}$\\
PO            & $1.4\times10^{-7}$     & $7\times10^{-8}$ \s{i}     & $-$     & (0.5--1) $\times$ $10^{-7}$ $^{\text{i}}$\\
NO            & $\sim$$10^{-9}$     & $-$     & $6.7\times10^{-6}$     & $-$\\
NS            & $2\times10^{-7}$     & $2\times10^{-9}$     & $1.7\times10^{-6}$     & $2.3\times10^{-8}$\\
NH$_3$ $^{\text{j}}$    & 2 $\times$ $10^{-6}$  & 5 $\times$$10^{-6}$    &  (2--5)  $\times$  $10^{-7}$ & $-$\\
AlO           & $4\times10^{-8}$     & $\sim$$6\times10^{-9}$     & $-$     & (4--8) $\times$ $10^{-8}$ $^{\text{l}}$ \\
AlOH          & $9\times10^{-8}$     & $-$     & $-$     & (2--4) $\times$ $10^{-9}$ $^{\text{l}}$\\
TiO $^{\text{k}}$     & $3\times10^{-9}$     & $-$     & $-$     & $\sim$$10^{-8}$\\
TiO$_2$ $^{\text{k}}$  & $3\times10^{-10}$     & $-$     & $-$     & $\sim10^{-8}$\\
VO            & $\sim$$10^{-9}$     & $-$     & $-$     & $-$\\
KCl           & $10^{-10}$     & $-$     & $-$     & $-$\\
AlCl          & $\sim$$10^{-8}$     & $-$     & $-$     & (0.1--2) $\times$ $10^{-8}$ $^{\text{l}}$\\
\bottomrule
 \end{tabularx}
\end{adjustwidth}
\noindent{\footnotesize{{\s{a}} Relative to H$_2$; $f$ = X/H$_2$; 
{\s{b}}~\cite{2009ApJ...695.1604Z, 2007Natur.447.1094Z, 2013ApJ...778...22A, Ravi2025a}; 
{\s{c}}~\cite{2022AJ....164..230S, 2021ApJ...920L..38S}; 
{\s{d}}~\cite{2016A&A...592A..51Q}; rotational diagram analysis; 
{\s{e}}~\cite{2009ApJ...695.1604Z, 2017A&A...597A..25V, 2020MNRAS.494.1323D, 2011ApJ...743...36P}; 
{\s{f}}~\cite{2021ApJ...907...42N}, SOFIA data; 
{\s{g}}~\cite{2004ApJ...610..427Z},  ISO, SWAS data; 
{\s{h}}~\cite{2011ApJ...743...36P}; 
{\s{i}}~\cite{2018ApJ...856..169Z}; 
{\s{j}}~\cite{2012A&A...545A..99T, 2016A&A...592A..51Q,1995ApJ...448..416M}
  {\s{k}}~\cite{2013A&A...551A.113K, 2020ApJ...904..110D}    };   {\s{l}}~\cite{2017A&A...598A..53D}.}
\end{table}

Nitrogen chemistry is important for IRC+10420. Note that N$_2$H$^+$ is only observed in this object. While NS is observed in all three stars, with an abundance range of \mbox{(0.002--1.7) $\times$ $10^{-6}$,}  it is most prevalent in IRC+10420. NO is also conspicuously present in this source, but only weakly visible in VY CMa.

Metal-bearing molecules and PO have not been observed thus far in IRC+10420 \cite{2016A&A...592A..51Q}. Comparisons can, therefore, be made only between VY CMa and NML Cyg. Here, the agreement between abundances is within a factor of two for NaCl and PO (Table~\ref{tab:abundances}). AlO is weaker in NML Cyg and its abundance is within a factor of $\sim$5--6 of that of VY CMa. NML Cyg does not quite have 
the variety of species that VY CMa has, but this result is likely a simple question of sensitivity. After longer integrations, KCl has been found in NML Cyg, \mbox{for example.}

Another comparison of note is RSGs vs. O-rich AGB stars. Representative AGB abundances are also listed in Table~\ref{tab:abundances}. There is a greater range in AGB values because more stars are compared. It is notable that the AGB abundance ranges typically span the values found in the RSG cases. This agreement is found in CO, HCN, CN, HCO$^+$, CS, SiO, NS, PO, PN and AlO. A few AGB values are on the high side relative to RSGs, in particular for SiS, H$_2$S, and NaCl, although the full range is consistent. 
  For example, higher abundances of $\sim$$10^{-7}$ are found for NaCl around AGB stars, while values for VY CMa and NML Cyg are $\sim$$10^{-9}$. These differences need further investigation.

\subsection{The Role of LTE Formation}
\label{sec:LTE}
\begin{figure}[H]
\begin{center}
\includegraphics[width=0.7\textwidth]{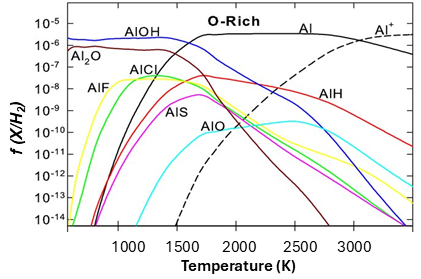}
  \end{center}
  \caption{{Results} 
 of model calculations of LTE abundances of Al-bearing molecules
relative to total hydrogen  (H$_2$ + H + H$^+$), plotted as a function of gas temperature, assuming
$\log P_{\mathrm g} = 3.0$, using the model of \cite{1973A&A....23..411T} with updated cosmic abundances \cite{2003ApJ...591.1220L}, and C/O$\sim$0.5.
    \label{fig:Mols9}}
\end{figure}

The high spatial resolution of the ALMA images allows a close examination of the gas-phase chemical processes occurring in VY CMa, and likely also in NML Cyg and other RSGs by inference. As predicted by Tsuji \cite{1973A&A....23..411T}, the temperatures and densities near the stellar photosphere in these stars ($\leq$$5 R_{\star}$) are sufficiently high ($T$$\sim$2000--3000 K; \mbox{$n$$\sim$$10^{10}$--$10^{11}$ cm$^{-3}$}) such that LTE conditions prevail. Three-body collisions occur and reaction barriers are easily overcome, such that the most stable species thermodynamically are produced. In this case, the chemical pathways are path-independent, and the abundances can be predicted with thermodynamic data, assuming a given temperature. A sample set of such calculations is shown in Figure~\ref{fig:Mols9}, from \cite{2009ApJ...694L..59T}, plotting fractional abundances {$f$} 
relative to hydrogen (H$_2$ + H + H$^+$), vs. temperature, for aluminum-bearing molecules. This plot could also be framed in terms of radius from the star ($\sim$1--8 $R_{\star}$ for $T_{\star}$$\sim$3000--3500 K, applicable to VY CMa \cite{2012A&A...540L..12W}). Here, the model from Tsuji \cite{1973A&A....23..411T} has been updated with revised cosmic abundances \cite{2003ApJ...591.1220L} and assumes oxygen-rich gas (C/O$\sim$0.5). As seen in the figure, near \mbox{T$\sim$3000 K}, aluminum is in the form of Al and Al$^+$. For VY CMa, this temperature is reached at about 1 $R_{\star}$.  As the gas cools to near 1000 K, AlOH is the dominant molecular species with $f$$\sim$$10^{-6}$. Al$_2$O is also important with a similar abundance, followed by AlF and AlCl ($5 \times 10^{-8}$). AlO peaks in abundance near 2500 K (1--2 $R_{\star}$), with $f$$\sim$$5 \times 10^{-10}$.

In comparison, observations towards VY CMa suggest that AlO and AlOH have similar abundances ($f$$\sim$$5 \times 10^{-8}$ and $9 \times 10^{-8}$, respectively) and AlCl has $f$$\sim$$10^{-8}$. However, all three Al-bearing species are localized near the star within about 50--60 $R_{\star}$, and disappear thereafter (see Figures~\ref{fig:Mols7} and~\ref{fig:Mols8}). Although the abundances do not match the model predictions exactly, the spatial extent of these molecules strongly suggests production by LTE chemistry. The higher observed abundance of AlO suggests that shocks may be disrupting corundum grains (Al$_2$O$_3$), producing an excess of this molecule relative to AlOH. As observed in Figures~\ref{fig:Mols7} and~\ref{fig:Mols8}, the images of PO also suggest LTE formation. In this case, LTE models predict a peak abundance of $\sim$$10^{-7}$, as is observed \cite{2024ApJ...971L..43R}, in excellent agreement. TiO$_2$ has a similar confined distribution  \cite{Ravi2025a}.
LTE chemistry, therefore, plays an important role in the synthesis of metal and phosphorus oxides in RSGs. 

In addition to gas-phase molecules, dust grains also form near the photosphere under LTE conditions \cite{2003ApJ...591.1220L, 2020A&A...637..A59}. Condensation is thought to be complete near 10--20 $R_{\star}$, although shocks can disrupt this process \cite{2006A&A...456.1001C}.  Because RSGs are oxygen-rich, materials that condense out are primarily oxides in nature, including amorphous silicates (SiO)$_{n}$, crystalline silicates such as spinel (MgAl$_2$O$_4$) and fosterite (Mg$_2$SiO$_4$), and corundum (Al$_2$O$_3$) \cite{1999IAUS..191..279L}. Phosphorus is thought to condense out as schreibersite, [Fe, Ni]$_3$P. Some of these materials can be observed as broad spectral features in the infrared, arising from lattice vibrations. For example, amorphous SiO has distinct features at 9.7 and 18 {\textmu}m attributed to the Si-O stretch in the grains, observed in NML Cyg \cite{1997ApJ...481..420M}. Other dust features are not as well characterized because of the ambiguity in solid-state spectra, which measure vibrational modes and not distinct compounds, unlike gas-phase data.  Grain composition is, therefore, often inferred from examining pre-solar grains extracted from meteorites \cite{2005ChEG...65...93L}. As a consequence, a quantitative assessment of grain composition is not possible. However, grain formation likely removes molecules such as AlO and PO from the gas phase and accounts for their disappearance beyond the inner envelope \cite{2024ApJ...971L..43R}.

\subsection{Influence of Shocks}
\label{sec:shocks}

\begin{figure}[H]
\begin{center} 
  \includegraphics[width=\textwidth]{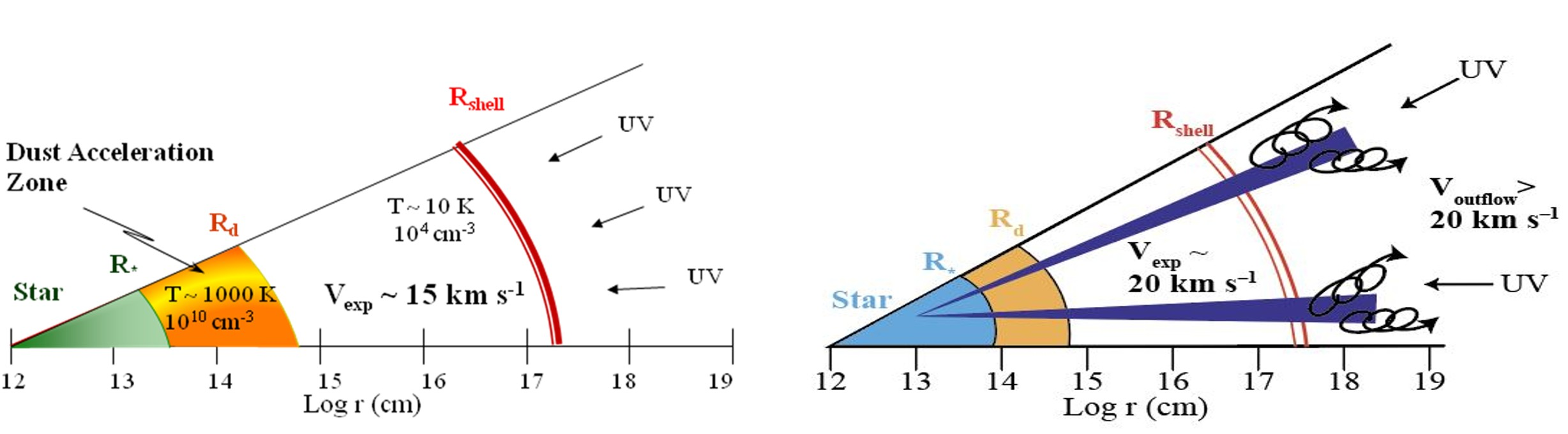}
  \vspace{-15pt}
    \end{center}
  \caption{{Models} 
 of circumstellar chemistry for AGB stars (\textbf{left}) and RSGs (\textbf{right}). Whilst AGB stars
have three basic chemical zones (LTE, “freeze-out” and photochemistry sectors), molecule formation in
RSGs is also strongly influenced by their highly directional, sporadic mass loss events that also disrupt
the basic envelope outflow.
    \label{fig:Mols10}}
\end{figure}

The chemistry in circumstellar envelopes is influenced by factors other than LTE synthesis. In the AGB scenario (see Figure~\ref{fig:Mols10}, {\textbf{{left}}}), LTE molecule and dust formation occur near the photosphere. At about 10--20 $R_{\star}$, all the dust has formed and the material reaches a terminal outflow velocity of 10--20 km s$^{-1}$. The outflow is sufficiently fast such that reaction timescales are too short to influence the overall composition, causing chemical ``freeze-out''. However, as the molecular material reaches the outer envelope ($\sim$500 $R_{\star}$), the density and temperature of the gas decrease considerably from that of the inner envelope. UV radiation from ambient starlight can, therefore, penetrate the envelope, initiating active photochemistry, creating a slew of free radicals and ions \cite{2006PNAS..10312274Z, 1996ARA&A..34..241G}; see Figure~\ref{fig:Mols10}. In addition, AGB stars pulsate, creating periodic shock waves in the inner envelope, which can alter molecular abundances \cite{2019A&A...629A.146V}. For example, the abundance of SiO is predicted to be negligible in lower mass C-rich AGB envelopes by LTE models \cite{2012A&A...545A..12C}; its common appearance in these objects is thought to arise from periodic shocks that increase the abundance by orders of magnitude near $\sim$1--5 $R_{\star}$ \cite{2012A&A...543A..48A}.

In principle, the envelopes of RSGs follow a similar morphology. There is LTE chemistry near the stellar photosphere, in gas from a lower velocity wind \cite{2013ApJ...778...22A} followed by transport through the middle to the outer envelope. Sharp temperature and density gradients are seen in RSG stars, and UV radiation and photochemistry may play a role in the outer regions. A typical temperature profile (in K) as a function of radius from the star, $r$ is \cite{2009ApJ...695.1604Z, 2004ApJ...610..427Z}:
\begin{linenomath}
\begin{equation}
  T_{\mathrm{gas}}\sim230 (10^{16}\mathrm{cm} /r)^{0.62}.
\end{equation}
\end{linenomath}
{However,} 
 it is clear from the HST and CO images of certain RSGs that any previous spherical shell is distorted. The asymmetric, energetic mass loss events characteristic of these objects disrupt the envelope, creating the knots and clumps seen in Figure~\ref{fig:Mols1}. These violent events must alter middle and outer envelope chemistry by creating localized shocks and entraining cooler and less dense material in faster outflows. These stark differences are illustrated in Figure~\ref{fig:Mols10}. The highly localized and non-spherical distributions of molecules, such as PN, NS and NaCl in VY CMa (Figures~\ref{fig:Mols7} and~\ref{fig:Mols8}) are good evidence for the influence of mass ejecta. The images of SiO, H$_2$S, and water in VY CMa closer in to the star are also highly anisotropic \cite{2023A&A...669A..56Q}.

 The shocks associated with RSGs differ in part from those in AGB envelopes. Although RSGs can undergo stellar pulsations, many of the shocks
 are thought to be associated with discrete mass loss events in which large cells of material are ejected from the stellar \mbox{surface \cite{2025AJ....169..230H}, }as distinct from periodic pulsations. They appear to be more energetic ($v_{\mathrm{outflow}}$$\sim$20--40 km s$^{-1}$) than shocks generated by pulsations which have $v \lesssim 20$ km s$^{-1}$ \cite{2012A&A...545A..12C}. RSG variability due to pulsations is often weak and irregular, and wind tracers close to the star, e.g., SiO masers have speeds $\lesssim15$ km s$^{-1}$, but with occasional, dramatic exceptions (Section \ref{sec6}). Violent ejections are individual, not recurring events, with timescales in the order of a few hundred years. Because they involve surface activity, they may contain material that is connectively mixed from the envelope. Low $^{12}$C/$^{13}$C ratios deduced from molecular lines observed in RSG envelopes are evidence of dredge-up from the H-burning shell \cite{2009ApJ...690..837M, 2023ApJ...954L...1S}; the surface activity may be related to additional mixing activities that penetrate beyond the H-burning shell. The chemistry in the envelope must be influenced by these unique shock and mixing processes, but a clear picture for RSGs has yet to emerge. Data are thus far limited.

 \subsection{Chemical Formation of Phosphorus Nitride: Mass Loss, Shocks, and Convective Mixing}\label{sec4.4}
 \label{sec:phosphors}

The best picture that can be formulated thus far for the unusual chemical processes in RSG envelopes is for VY CMa, based on the large set of ALMA images, focusing on the well-analyzed phosphorus system \cite{2024ApJ...971L..43R}. It is clear that PO is solely confined to the inner circumstellar envelope ($<$60 $R_{\star}$) and must be primarily a product of LTE chemistry (Figure~\ref{fig:Mols8}). While PN has a similar inner shell distribution, it is prominent in other structures 80--240 $R_{\star}$ from the star, i.e., in the SW Fan, some of the SW Clumps, and in the four Cloudlets, as mentioned (Figure~\ref{fig:Mols8}). These distances are well beyond the LTE chemical region.  The velocities of the PN spectra in the SW Clump region are close to those of K{\sc i} present in several of the actual S and SW knots (S-D, SW-A, SW-G, S-E1/E2 \cite{2024ApJ...971L..43R, 2025AJ....169..230H}). The presence of PN in these structures is most likely related to the mass ejections that created them, which occurred on short timescales about 200--300 years ago. These ejecta must represent highly shocked material, considering the clump velocities \cite{2025AJ....169..230H}, and may have caused phosphorus to be liberated from dust grains, perhaps composed of schreibersite.  Evidence for the shock scenario is the similarity between PN and SiO emission in the SW Fan (see Figure~\ref{fig:Mols8}). SiO is also thought to be associated with shock-initiated chemistry \cite{1989ApJ...343..201Z}. The presence of PN in Cloudlets I--IV may  have a shock origin, as well, as these structures could have been generated in the same event that created the S and SW clumps.  The Cloudlets lie about the same distance from the star, but are obscured by extinction in HST images.

The striking result is that PN is found in material where PO is negligible. PO and PN are almost always observed together, and shock models predict their joint \mbox{formation \cite{2012ApJ...761...74A, 2018ApJ...862..128J, 2016MNRAS.462.3937L}. } However, models predict that the PO/PN ratio is sensitive to the nitrogen abundance, and enhancements of this element result in a significantly higher abundance of PN relative to PO by several orders of magnitude. The chemical time scales need to be short \mbox{($\sim$100--200 years)}, however, or PO forms. If the clump ejecta involve material that is convectively mixed from the H-burning shell, they may be enriched in nitrogen. In the H-burning shell, N enhancement usually occurs with excess $^{13}$C production, lowering the $^{12}$C/$^{13}$C\mbox{ ratio \cite{1973ApJ...184..493A}}.  Estimates of this ratio in the SW Fan suggest $^{12}$C/$^{13}$C$\sim$7, based on HCN lines \cite{2023ApJ...954L...1S}, consistent with this scenario. Furthermore, as shown in Figure~\ref{fig:Mols8}, there is a remarkable correlation between PN and NS in the ALMA images, suggesting that both molecules arise in nitrogen-rich gas. These data show the influence on RSG circumstellar chemistry of the discrete mass loss and associated shocks, as well as that of possible convective mixing in the ejection processes.  Note that in some AGB envelopes there are localized concentrations of NS and SiC, for example, but these structures appear to be attributable
 to wind disruption and shocks arising from the interaction of a companion star \cite{2024NatAs...8..308D, 2025MNRAS.536..684D}.

\section{Molecules, Isotopes and Nucleosynthesis}
\label{sec:isotopes}
Molecular lines offer another avenue of insight into the envelopes of RSGs: isotope ratios. These ratios provide a window into the nucleosynthesis occurring in the shell-burning layers of the star and the degree of convection. For RSGs, the convective layer extends to deep within the envelope and should mix products of nucleosynthesis to the surface \cite{2017ars..book.....L}. The most likely products are those from the H-burning shell.

Table~\ref{tab:isotopes} lists the isotope ratios measured in the envelopes of RSG stars. Unfortunately, the number of values is very limited. The dearth of ratios is in part due to the complexity of many of the RSG envelopes, such as that of VY CMa. It is difficult to measure accurate ratios in multiple overlapping regions \cite{2023ApJ...954L...1S, 2022AJ....164..230S, 2021ApJ...920L..38S}. As Table~\ref{tab:isotopes} shows, the $^{12}$C/$^{13}$C ratio has been measured in two RSGs VY CMa and NML Cyg. The values from \cite{2009ApJ...690..837M, 2022AJ....164..230S, 2021ApJ...920L..38S, Ravi2025a} 
were measured from abundances derived from radiative transfer modeling of spectral lines, while the others were estimated solely from line intensities \cite{2023ApJ...954L...1S, 2022MNRAS.510..383A}.  The values have been determined from CO and HCN, respectively, both of which may be optically thick in the main isotopologue. Also, the data used were from ALMA \cite{2023ApJ...954L...1S, Ravi2025a}
and single-dish observations \cite{2009ApJ...690..837M, 2021ApJ...920L..38S, 2022MNRAS.510..383A}.  As shown in Table~\ref{tab:isotopes}, the ratio in NML Cyg is $33 \pm 15$ (ARO SMT), while that for VY CMa falls in the range $^{12}$C/$^{13}$C $\sim$ 22--38 (ALMA) and 26--46 (ARO SMT). The range of values in the latter source reflects the various regions observed. From the ALMA data, the highest value of 38 is found in Arc 2  (see Figure~\ref{fig:Mols1}).  The lowest ratio of 22 is measured in the SW Clump, which consists of three velocity components at 19,  --11, and 50 km s$^{-1}$ with corresponding ratios of $^{12}$C/$^{13}$C = 22, 26, and 33, respectively. The NW Arc, NE Arc, and NE Extension exhibit values in the range 23--30. This scatter in the ratios may represent different mass loss events with varying degrees of mixing \cite{2009ApJ...690..837M, 2023ApJ...954L...1S}, or they may be more indicative of the uncertainties. Nonetheless, these  values are low relative to the solar value of $^{12}$C/$^{13}$C = 89, and more importantly, to the envelopes of C-rich AGB stars, which is $^{12}$C/$^{13}$C$\sim$25--90 \cite{2009ApJ...690..837M}. The lower values in the RSGs reflect the mixing into the H-burning shell. For AGB stars, a third dredge-up occurs, where material rich in $^{12}$C is pulled up from the He-burning shell \cite{2009ApJ...690..837M}. The third dredge-up on the AGB increases the $^{12}$C/$^{13}$C ratio. The carbon ratios measured in VY CMa and NML Cyg thus agree with current models of nucleosynthesis. Note that Red Giant Stars (RGSs) also have low ratios (see Table \ref{tab:isotopes}).

The three oxygen isotopes are indicators of the relative amount of dredged material in H vs. He-burning shells\cite{2016PrPNP..89...56B}. Unfortunately, only the $^{16}$O/$^{18}$O ratio has been measured thus far in RSGs, solely in NML Cyg. The ratio is not well-determined at \mbox{$^{16}$O/$^{18}$O $>$ 250 \cite{2021ApJ...920L..38S},} established from SiO abundances.  Another value derived from water lines observed with \textit{{Herschel}}/HIFI, $8.7 \pm 0.5$, is quite low and would imply an extraordinary enhancement of \mbox{$^{18}$O \cite{2022MNRAS.510..383A}.} Both $^{16}$O and $^{18}$O are produced primarily in He-burning,  the former from the reaction of $^{12}$C and an alpha particle ($^4$He), and the latter from $^{14}$N + $^4$He \cite{2016PrPNP..89...56B}. Ne-burning creates additional $^{16}$O. It would seem unlikely that these processes would result in such excess $^{18}$O in RSGs. The low ratio is likely a result of high opacity in the H$_2$O lines of the main isotopologue. The data, as well as that in RGSs (Table \ref{tab:isotopes}), therefore, imply a relative agreement with the solar value of 500. Additional measurements of oxygen isotopes \mbox{are needed.}

\begin{table}[H]
\caption{ Measurements of Isotope Ratios in RSG Envelopes. \label{tab:isotopes}}

\small
\begin{adjustwidth}{-\extralength}{0cm}
\begin{tabularx}{\fulllength}{LLLLLLLL}
\toprule
&\boldmath{$^{12}$}\textbf{C/}\boldmath{$^{13}$}\textbf{C}&\boldmath{$^{16}$}\textbf{O/}\boldmath{$^{18}$}\textbf{O}& \boldmath{$^{35}$}\textbf{Cl/}\boldmath{$^{37}$}\textbf{Cl}& \boldmath{$^{32}$}\textbf{S/}\boldmath{$^{34}$}\textbf{S}& \boldmath{$^{28}$}\textbf{Si/}\boldmath{$^{29}$}\textbf{Si}& \boldmath{$^{28}$}\textbf{Si/}\boldmath{$^{30}$}\textbf{Si}&\boldmath{$^{29}$}\textbf{Si/}\boldmath{$^{30}$}\textbf{Si}\\
\midrule
VY CMa \s{a} & 22--38 \s{f}\linebreak 25--46 \s{
f} &       -- & $3.6 \pm 0.6$&  --  &  -- &  -- &  -- \\
\midrule
NML Cyg \s{b}&$33 \pm 15$; 17; $9.3 \pm 1.5$ & $>250$; $8.7 \pm 0.5$ &$3.9 \pm 0.9$&$50 \pm 25$&$33 \pm 10$; $10.9\pm 0.9$&$33 \pm 10$; $9.5 \pm0.7$& $1.2\pm 1.0$; $0.9 \pm 0.4$\\
\midrule
VX Sgr \s{c}&  --  &  -- &  -- &  -- &  --  &  -- & $4\pm 1$\\
\midrule
RGS \s{d}&  7-17 $\pm$ 4 &  550 $\pm$ 150 &  -- &  -- &  --  &  -- &  --\\
\midrule
Solar  \s{e} & 89 & 500 & 3.13 & 25 & 20 & 29 & 1.5 \\
\bottomrule
\end{tabularx}
\end{adjustwidth}
\noindent{\footnotesize{\s{a} \cite{2023ApJ...954L...1S} from ALMA; \cite{2009ApJ...690..837M} from ARO SMT ; 
\s{b} \cite{2021ApJ...920L..38S} ARO SMT, \cite{2009ApJ...690..837M} ARO SMT, \cite{2022MNRAS.510..383A} OSO; 
\s{c} \cite{2024A&A...681A..50W}; 
\s {d} \cite{1984ApJ...285..674H}; 
\s{e} \cite{2009ARA&A..47..481A}; 
\s{f} Range reflects differing regions}.}
\end{table}

The chlorine isotope ratio has been measured in both VY CMa and NML Cyg, with values of $^{35}$Cl/$^{37}$Cl = $3.6 \pm 0.6$ and $3.9 \pm 0.9$. The ratios were determined from NaCl with ALMA  \cite{Ravi2025a}
and ARO SMT \cite{2021ApJ...920L..38S} observations. These ratios agree well with the solar value of 3.13.  Both the Cl isotopes are formed primarily in hydrostatic (shell) and explosive oxygen burning, as well as Ne-burning for $^{37}$Cl. The ratios measured, therefore, likely reflect previous supernova explosions and subsequent mixing.

Silicon isotope ratios were measured from SiO in NML Cyg; those established from radiative transfer analyses are consistent with the solar ratios, with $^{28}$Si/$^{29}$Si$\sim$$33 \pm 10$, $^{28}$Si/$^{30}$Si$\sim$$33 \pm 10$, and $^{29}$Si/$^{30}$Si$\sim$$1.2 \pm 1.0$ \cite{2021ApJ...920L..38S}. The solar values are 20, 30, and 1.5, respectively \cite{2009ARA&A..47..481A}. Much lower values were found from comparing line intensities (Table~\ref{tab:isotopes}), but are probably due to high opacities in $^{28}$SiO. Agreement with solar values is not unexpected, as $^{28}$Si is formed in O-burning, and $^{29}$Si and $^{30}$Si in Ne-burning \cite{2016PrPNP..89...56B}. These shells are unlikely to be accessed by convection in the RSG phase. The same is found for the $^{32}$S/$^{34}$S ratio. As measured from SiS \cite{2021ApJ...920L..38S}, the ratio is $50 \pm 25$ in NML Cyg, with a solar value of 25---comparable within the uncertainties. Again, the sulfur isotopes are a product of oxygen burning \cite{2016PrPNP..89...56B}. 

\section{Molecular Masers in RSGs}
\label{sec6}
Molecules in RSGs are often found to exhibit maser action in certain species. Cosmic masers are the naturally occurring analogs of lasers, which occur in molecular rotational transitions from MHz to THz frequencies. Masers occur when molecules are radiatively or collisionally excited (``pumped'') to a higher energy level,  which then decays into lower energy states. At a certain level, the cascading radiation reaches a “bottleneck,” such that the population builds. The level then becomes overpopulated with respect to the next level below, causing “population inversion.”. Maser amplification occurs when a background photon triggers an exponential decay of the built-up population to the lower level. Often, the ambient radiation is amplified, but seed photons can come from bright backgrounds such as the star.   
  {Population inversion occurs under a specific range of temperatures, number densities, abundances, radiation field and velocity gradients for each masing transition.}

\begin{figure}[H]
\begin{center}
  \includegraphics[width=0.4\textwidth]{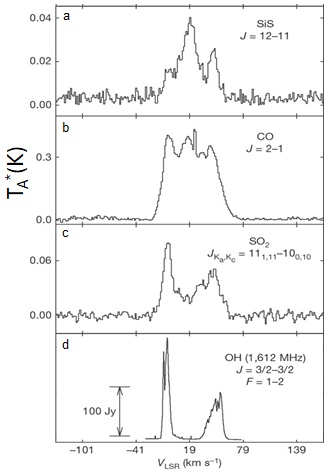}
  \end{center}
     \caption{{Spectra} 
 of the OH maser line(1612 MHz; (\textbf{d})) and thermal lines of SO$_2$, CO, and SiS ((\textbf{a})--(\textbf{c})) from \cite{2007Natur.447.1094Z}. The OH lines are narrower and occur at distinct velocities, while the thermal lines cover a broad, continuous velocity range. }
    \label{fig:Figure12}  
\end{figure}
\begin{figure}[H]
\begin{center}
\includegraphics[width=0.8\textwidth]{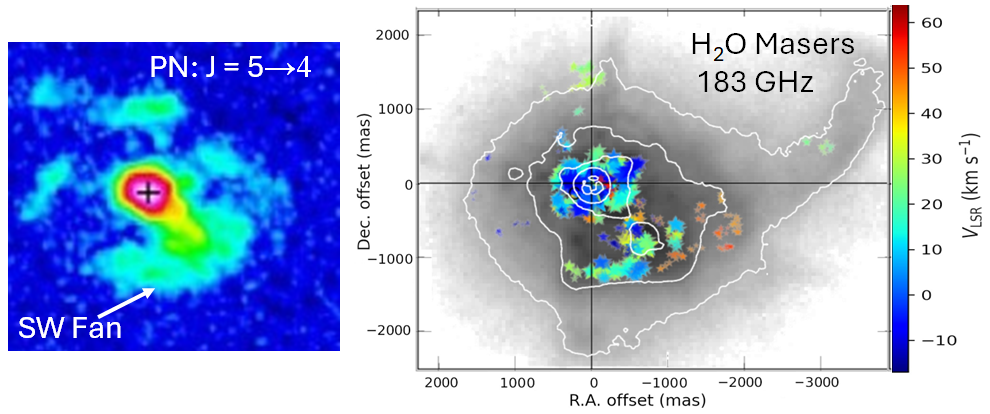}
\end{center}
     \caption{({\textbf{Right}}): {The} 
       distribution of H$_2$O 183 GHz masers in the envelope of VY CMa (colour scale in velocity) overlaid on an HST/WFPC2 F1042M image \cite{2017ApJ...851...65D, 2001AJ....121.1111S}. $R_{\star}$ is 5.7 mas \cite{2012A&A...540L..12W}. (\textbf{{Left}}): PN image from Figure~\ref{fig:Mols8} \cite{2024ApJ...971L..43R} plotted on an identical scale. The cross marks the position of the star. The water masers appear to trace
       the shock front associated with the SW Fan, as traced in PN.}
    \label{fig:Figure13}  
\end{figure}

{The} exponential nature of maser amplification implies that, spectrally, the line center is amplified more than the wings and so the line shape becomes narrower relative to thermally-excited transitions (e.g., Figure \ref{fig:Figure12}  ). Because of the amplification, maser emission lines are also quite intense, and can reach hundreds to thousands of Jy, with brightness temperatures exceeding $10^{12}$ K and narrow linewidths $<1$ km s$^{-1}$. The longer the path length of gas with a population inversion in a given velocity interval, the greater the amplification, so maser spots can appear much smaller than the emitting cloud -- “maser beaming”. 
Note that small changes in underlying conditions can produce dramatic changes in maser appearance.
Masers can be excited by shock compression and heating, particularly effectively
  perpendicular to the flow direction, providing a useful shock diagnostic \citep{Elitzur92a, Richards11}.
  Excellent examples of this phenomenon are seen in VY CMa, to the E \citep{2014A&A...572L...9R}, in the bowshock-like 268-GHz maser towards the NW outflow \citep{2023atyp.confE..72R} and possibly as shown in Figure \ref{fig:Figure13}, where a line of H$_2$O 183 GHz maser spots appears to trace a shock front observed in the PN molecule, which deliniates the SW Fan (see Section \ref{sec4.4}).
    More details about interstellar and circumstellar masers can be found in \citet{Gray2012},  \citet{Elitzur1992}, and \citet{Cohen1989}.

\vspace{-12pt}

The position and size of maser spots can be measured to an accuracy proportional to the (synthesized beam)/(signal-to-noise ratio), within the uncertainties in these two \mbox{quantities  \cite{1997PASP..109..166C, Richards11}.}
This allows the location and kinematics of the masing clumps to be measured with precision typically an order of magnitude greater than thermal (non-maser) observations. Proper motions of maser spots {can be measured} 
in months or even weeks, and such “maser parallax” measurements can be used to accurately measure distances to stars and other objects even more reliably than {\it{GAIA}}. For example, maser parallax determinations have established distances for VX Sgr at $1.56^{+0.11}_{-0.10}$ kpc \citep{2018ApJ...859...14X} and VY CMa at $1.20^{+0.13}_{-0.10}$ kpc \citep{Zhang12} or $1.14^{+0.11}_{0.99}$ kpc \cite{Choi08}. 


\subsection{Types of Masers in RSGs}
\label{sec:maserstudies}

{The masers} that occur in RSGs usually arise from SiO, H$_2$O and OH. SiO maser excitation appears complex, including collisional and radiative pumping and line overlaps \cite{2002A&A...386..256H}.  Most H$_2$O transitions are usually collisionally excited \cite{2016MNRAS.456..374G}. Ground-state OH masers, arising from  hyperfine components \textit{F} within a lambda doublet, are radiatively excited by warm dust in CSEs \cite{1994ApJ...422..193C, 1976ApJ...205..384E}. OH 1612-MHz emission varies in response to the stellar variability. The time-lag due to light travel time between the lower (near-side) and higher (far side) velocity peaks gives the depth of the OH shell and this can be compared with its angular size to give the `phase-lag' distance, e.g. \cite{2018IAUS..336..381E}.
 
  SiO masers typically occur in rotational transitions in excited vibrational states (v $>$ 0)  in the $J = 1 \rightarrow 0$ rotational transitions near 43 GHz,  as well as in higher rotational states \cite{2009MNRAS.394...51G}, and have been detected up to $J = 15\rightarrow 14$ at 646 GHz \cite{2018A&A...609A..25B}.    
   These masers (energy levels) $\ge$ 1800 K) are found in spots typically within a few $R_{\star}$, forming a whole or partial ring around the star. They trace  parts of an accelerating (or decelerating)  shell, which favors maser amplification from material in the plane of the sky (``tangential beaming'', first recognized in OH masers \cite{1985MNRAS.212..375C}).
 Monitoring and proper motion studies of SiO maser spots \cite{2002AAS...20111509O, Zhang12, 2024AJ....168...53Y} show
infall as well as outflow, tracing the competition between gravity and shocks from stellar pulsation and mass loss events.  This competition means that the maser material spends longer in a given region than predicted from a simple linear outflow\cite{2018ApJ...869...80A}, giving the material more time for dust formation and other chemistry. As strong molecular tracers, SiO masers are least likely to be disrupted by interstellar interactions. They, therefore, help to identify RSGs in globular and other clusters, e.g., \cite{2005PASJ...57L...1M, 2012A&A...541A..36V}.

There are over 50 potentially observable   H$_2$O maser transitions, ranging in energy levels from 200 to $>$6000 K, each with distinctive pumping conditions. 
About two dozen different water masers have been detected since the 22 GHz line $J_{\mathrm{Ka,Kc}}  = 6_{1,6}\rightarrow5_{2,3}$  discovery \citep{1969Natur.221..626C}, extending up to 1.885 THz \citep{2017ApJ...843...94N}.
The highest-energy lines \mbox{(states $>$ 3000 K)} are mostly confined to the inner few $R_{\star}$,  e.g., \cite{2023A&A...674A.125B}.
The inner radius of the region around the star exhibiting the 22 GHz maser is determined by a collisional quenching effect
at \mbox{$T_{\mathrm k}$$\sim$1000 K} for  $n\gtrsim 5 \times10^{9}$ cm$^{-3}$, suggesting that the masers arise in regions of dense clumps  at least an order of magnitude higher than the outflow average \cite{Richards12}. 
The lowest-excitation H$_2$O maser at 183 GHz ($J_{\mathrm{Ka,Kc}}= 3_{1,3}\rightarrow2_{2,0}$)
has a high detection rate in RSGs \citep{1998A&A...334.1016G, 2017A&A...603A..77H}, including sources such as $\mu$ Cep, NML Cyg, S Per, VX Sgr and VY CMa. This maser has broader lines than other water masers, and is located in circumstellar envelopes close to terminal velocity, near 10--20 $R_{\star}$. Figure \ref{fig:Figure14} shows the typical location of the water masers in VY CMa as a function of distance from the star. 

The most common OH  masers arise from hyperfine splitting in  the ground state lambda doublet ( $J = 3/2; \Omega = 3/2$) \cite{Gray2012}.
The main lines at 1665 and 1667 MHz are favored at
$T_{\mathrm k}\lessapprox500$ K and $n\lessapprox10^{8}$ cm$^{-3}$.
The 1612 MHz satellite line requires even cooler conditions and long (typically $\gtrsim$10 au) velocity-coherent paths, and is usually found at hundreds $R_{\star}$. Satellite line emission brighter than the mainlines indicates a CSE old enough to have built up a thick, stable outer shell at hundreds of stellar radii or more from the star.
{The only thermal OH detected from CSEs is in much higher excitation states, e.g., \citep{2019A&A...623L...1K, 2023A&A...674A.125B}.}

\subsection{Masers Tracing Wind Conditions}

\begin{figure}[H]
\begin{center}
  \includegraphics[width=0.6\textwidth]{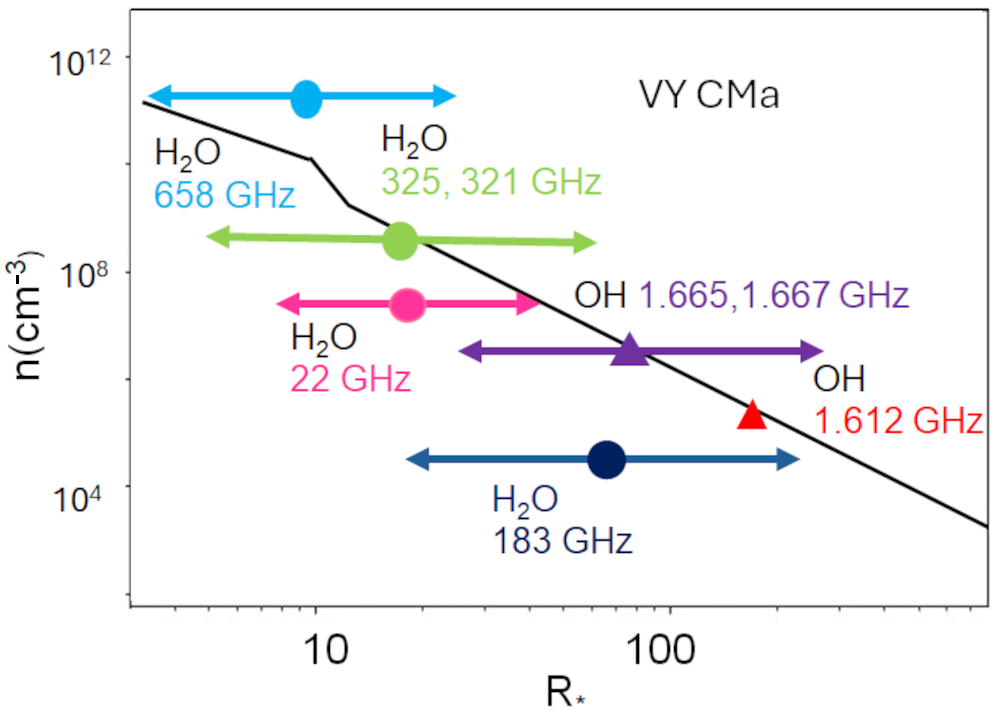}
  \end{center}
     \caption{{Qualitative} 
       plot showing the positions of OH and H$_2$O masers in VY CMa, labeled by their frequency (in GHz), and typical number densities for excitation \cite{Gray2016}.  The black line shows the approximate density profile,  modeled from Herschel CO data \cite{2006A&A...456..549D}. 
 The predicted -- and observed -- locations \cite{Cohen1989, Gray2016, 2014A&A...572L...9R,  2017ban5.conf...19R}  of masers are labeled by their frequencies (GHz).}
    \label{fig:Figure14}  
\end{figure}

Different types of masers occur at increasing distances from the star, as illustrated qualitatively in Figure \ref{fig:Figure14}. This is not a rigid sequence. Different pumping requirements (e.g., lower or higher number densities) make masers excellent tracers of inhomogeneous conditions at similar radii. The sequence of maser zones with distance from the star is
\noindent similar to that seen in AGB stars. The main difference is approximately an order of magnitude in spatial scale, due to the greater mass loss rates and extent of RSGs and their envelopes (and thus more gradual temperature and density gradients). The extent of masing regions also  scales with  the optical stellar radius $R_{\star}$  \cite{Richards12}. The number and size of such clumps suggests a link to giant convection cells; however, RSGs have more evidence for additional, sporadic, massive ejecta than lower mass stars (Sections~\ref{sec:shocks}).

{Water} transitions other than 22 GHz have only been mapped in a few RSG, e.g., \cite{Hunter07, 2014A&A...572L...9R, Asaki2020, 2023A&A...674A.125B, 2023atyp.confE..72R}, mostly around VY CMa, Figure \ref{fig:Figure14}. The 268, 658, 321, 22, 325 and 183 GHz masers usually extend to increasing radii with decreasing excitation,  tracking the wind from dust formation to acceleration through the escape velocity at tens $R_{\star}$, e.g., \cite{2014A&A...572L...9R}. Models for pumping and excitation provide constraints on the local number density $n$ and kinetic temperature \linebreak $T_{\mathrm k}$ \citep{Gray2016}. Figure \ref{fig:Figure14} shows that the radial gradient of $n$ is broadly consistent with the model of \cite{2006A&A...456..549D}, if most of the gas supporting H$_2$O masers is  concentrated in dense clumps  \cite{2023atyp.confE..72R}. Most species locations are also consistent with the  $T_{\mathrm K} (r)$ gradient in the envelopes, apart from  some exceptional 268 and 658 GHz masers from VY CMa, probably shock-associated.

 Masers from RSG at kpc distances can be at least as bright (sometimes exceeding thousands Jy) as much closer AGB masers, but there is no clear-cut relationship between masers and stellar type. The thermal line width and sound speed (affecting the velocity-coherent amplification depth) are independent of clump size. Even small clumps can produce maser flares if they overlap or are excited by other mechanisms such as \mbox{shocks \cite{2018evn..confE..73G}.}

Maser lines can also be used to measure magnetic fields in RSG envelopes through the Zeeman effect. For example, interaction between an external magnetic field and the OH radical's unpaired electron produces strong Zeeman splitting, e.g., 1.2 km s$^{-1}$ for a 2 mG field along the line of sight for the 1665 MHz line. 
The field strength and direction can be retrieved by modeling circular and linear polarization, \cite{1998ApJ...504..390E, Gray2012}, e.g. falling from 500 mG (22-GHz H$_2$O) at a few tens $R_{\star}$ to 3 mG at approaching 1000  $R_{\star}$ (1612 MHz OH)  around NML Cyg \cite{2002A&A...394..589V, 2004MNRAS.348...34E}.
Faraday rotation of linear polarization vectors by a few degrees  is seen within  a few $R_{\star}$ {e.g.} 
\cite{2004evn..conf..209R}
 implying an ionization fraction $\sim$$10^{-6}$, assuming $n$$\sim$$5\times10^{7}$ cm$^{-3}$.

The other masing molecules, such as water and SiO, do not have such a prominent Zeeman effect, as they have no unpaired electrons. The magnetic fields required for splitting the lines of SiO masers can reach 10--1000 G within a few $R_{\star}$ (e.g., $\mu$ Cep, \citep{2024A&A...688A.143M}). Comparing Zeeman results for masers excited at increasing distances from the star, the field strength appears to vary as $\approx$$r^{-2}$ \citep{2011ApJ...728..149V} and to have sufficient energy density to shape but not drive the wind.  There are exceptions depending on local density variations  \cite{2004evn..conf..209R}.

\section{Prospectus}

The past decade of observations has clearly demonstrated that the envelopes of RSGs have a unique and complex chemistry. The prominent examples are VY CMa and NML Cyg. Refractory oxides, such as PO, AlO, TiO, AlOH and VO highlight their unusual molecular inventory. More interesting molecules are likely to be discovered in these objects. Both RSGs have mass loss that is highly sporadic and directional, creating complicated structures not seen in AGB envelopes. The role of shocks in molecule formation is, therefore, critical and has already been observed in VY CMa through observations with ALMA. Molecule formation under LTE conditions is also important near the stellar photosphere. Not all envelopes of RSGs exhibit the complex structures and molecular differentiation that is seen in VY CMa and NML Cyg,  although some, like the YHG IRC+10420, have a similar mass loss rate. Here, there is likely an age factor at play, as YHGs are more evolved than Red Hypergiants. They also appear to have more detached, and therefore, thinner shells. This fact may explain the absence of many inner shell molecules such as metal halides and oxides.  The chemistry and molecular content of RSGs have only begun to be explored. As more observations are conducted, particularly with high spatial resolution and sensitivity, as offered by ALMA, many of these questions will be addressed.

\vspace{+6pt}

\authorcontributions: {The authors contributed equally to the planning, writing, and editing the manuscript.} 

\funding 
{This research is supported by NSF Grant AST-2307305, and NASA grants 80NSSC18K0584 (Emerging Worlds) and 80NSSC21K0593 for the program “Alien Earths".}
\acknowledgments 
 {The authors acknowledge the staff of the Arizona Radio Observatory for many years of successful observations with the ARO SMT and 12 m telescopes.
The following ALMA data were used: ADS/JAO.ALMA 2001.0.00011.SV, ADS/JAO.ALMA 2019.1.00659.L, ADS/JAO.ALMA 2021.1.01054.S. ALMA is a partnership of ESO, NSF, and NINS, with NRC, MOST, ASIAA, and KASI, in cooperation with the Republic of Chile. The Joint ALMA Observatory is operated by ESO, AUI/NRAO and NAOJ. NRAO is a facility of the NSF operated under cooperative agreement by Associated Universities, Inc.  MERLIN is a National Facility operated by the University of Manchester at Jodrell Bank Observatory on behalf of STFC.   The VLBA is operated by the National Radio Astronomy Observatory, a facility of the National Science Foundation operated under cooperative agreement by Associated Universities, Inc. This research has made use of NASA's Astrophysics Data System Bibliographic Services. }

\conflictsofinterest {The authors declare no conflict of interest.} 

\abbreviations{Abbreviations}{
  The following abbreviations are used in this manuscript:\\
  
  \noindent
\begin{tabular}{@{}ll}    
AGB           & Asymptotic Giant Branch\\ 
ALMA          & Atacama Large Millimeter/sub-millimetre Array\\
ARO SMT       & Arizona Observatory Sub-Millimeter Telescope\\
CSE           & Circumstellar Envelope \\
E-AGB         & Early-AGB\\
HST           & Hubble Space Telescope\\
IRAM          & Institute de Radioastronomie Millimetrique\\
ISM           & Interstellar Medium\\
JBCA          & Jodrell Bank Centre for Astrophysics \\
LSR         & Local Standard of Rest (kinematic) velocity frame\\
LTE           & Local Thermodynamic Equillibrium\\
MERLIN        & Multi Element Radio Interferometry Network\\
$r$          & Radial separation from star\\
$P_{\mathrm g}$& Gas pressure \protect{\cite{1973A&A....23..411T}}\\
$R_{\star}$    & Stellar radius (optical/IR)\\
RGB           & Red Giant Branch\\
RSG           & Red Supergiant\\
SMA           & Sub-Millimetre Array\\
\end{tabular}

\noindent
\begin{tabular}{@{}ll}

SNe           & \hspace{+9pt}Supernovae\\
$T_{\mathrm g}$ & \hspace{+9pt}Gas Temperature\\
$T_{\mathrm k}$ & \hspace{+9pt}Kinetic Temperature\\
$T_{\star}$    & \hspace{+9pt}Stellar temperature\\
TP-AGB        & \hspace{+9pt}Thermal Pulse-AGB\\
U-lines      & \hspace{+9pt}Unidentified lines\\
VLBA         & \hspace{+9pt}Very Long Baseline Array\\
VLBA         & \hspace{+9pt}Very Long Baseline Interferometry\\
$v$         & \hspace{+9pt}Stellar wind expansion velocity or vibrational state\\
$v_{\mathrm{outflow}}$ & \hspace{+9pt}Stellar wind velocity\\
YHG         & \hspace{+9pt}Yellow Hypergiant (RSG)\\
\end{tabular}
}



\begin{adjustwidth}{-\extralength}{0cm}
\reftitle{References}

\PublishersNote{}
\end{adjustwidth}
\end{document}